\documentclass[12pt,iopams]{iopart}

\usepackage{graphicx}

\newcommand{\bea}{\begin{eqnarray}}
\newcommand{\eea}{\end{eqnarray}}
\newcommand{\beq}{\begin{equation}}
\newcommand{\eeq}{\end{equation}}

\newcommand{\tseref}[1]{(\ref{#1})}

\newcommand{\keywords}[1]{\noindent{\emph{Keywords}\/} #1}

\newcommand{\tsepreprint}[1]{#1}

\newcommand{\tsetitle}[1]{\emph{#1}}

\newcommand{\tsemat}[1]{{\mathbf{\textsf{#1}}}}

\newcommand{\Amat}{\tsemat{A}}

\newcommand{\unitmat}{\hbox{\textsf{1}\kern-.25em{\textsf{I}}}}

\newcommand{\An}{A^{(n)}}
\newcommand{\Athree}{A^{(3)}}
\newcommand{\Amax}{A^{(\mathrm{max})}}
\newcommand{\Amaxthree}{A^{(\mathrm{max},3)}}
\newcommand{\Amaxfour}{A^{(\mathrm{max},4)}}
\newcommand{\Amaxn}{A^{(\mathrm{max},n)}}

\newcommand{\Bn}{B^{(n)}}
\newcommand{\Bthree}{B^{(3)}}

\newcommand{\Bbar}{\bar{B}}

\newcommand{\Cn}{C^{(n)}}
\newcommand{\Cthree}{C^{(3)}}
\newcommand{\Cfour}{C^{(4)}}
\newcommand{\Cfive}{C^{(5)}}

\newcommand{\Cmaxn}{C^{(\mathrm{max},n)}}

\newcommand{\Dn}{D^{(n)}}
\newcommand{\Dtilden}{\tilde{D}^{(n)}}
\newcommand{\Dtwo}{D^{(2)}}
\newcommand{\Dthree}{D^{(3)}}
\newcommand{\Dfour}{D^{(4)}}
\newcommand{\Dfive}{D^{(5)}}
\newcommand{\Dtildethree}{\tilde{D}^{(3)}}
\newcommand{\Dtilde}{\tilde{D}}

\newcommand{\kn}{k^{(n)}}

\newcommand{\Ln}{L^{(n)}}

\newcommand{\Pn}{P^{(n)}}
\newcommand{\Pthree}{P^{(3)}}

\newcommand{\str}{s}

\newcommand{\Ccal}{\mathcal{C}}
\newcommand{\Ccalthree}{\mathcal{C}^{(3)}}
\newcommand{\Ccaln}{\mathcal{C}^{(n)}}

\newcommand{\Ccalmaxthree}{\mathcal{C}^{(\mathrm{max},3)}}
\newcommand{\Ccalmaxfour}{\mathcal{C}^{(\mathrm{max},4)}}
\newcommand{\Ccalmaxn}{\mathcal{C}^{(\mathrm{max},n)}}

\newcommand{\Vcal}{\mathcal{V}}

\begin{document}


\title{Clique Graphs and Overlapping Communities}
\author{T.S.\ Evans}
\address{Theoretical Physics, and the Institute for Mathematical Sciences,
Imperial College London, SW7 2AZ, U.K.}

\date{9th September, 2010}

\begin{abstract}
It is shown how to construct a clique graph in which properties
of cliques of a fixed order in a given graph are represented by
vertices in a weighted graph.  Various definitions and
motivations for these weights are given.  The detection of
communities or clusters is used to illustrate how a clique
graph may be exploited. In particular a benchmark network is
shown where clique graphs find the overlapping communities
accurately while vertex partition methods fail.
\end{abstract}




\submitto{Journal of Statistical Mechanics}

\maketitle

\renewcommand{\thefootnote}{\arabic{footnote}}
\setcounter{footnote}{0}

\section{Introduction}\label{sintro}

Much of the work on networks is from a vertex centric viewpoint.
We talk about distributions of vertex degree, the clustering
coefficient of vertices, and vertex partitions as communities. For
instance, consider table \ref{tnetrev} which shows the frequency of words in a review of
networks \cite{Evans04}.  If we ignore stop words such as ``the''
and ``a'', and use the stems of words (so `edg' represents ``edge'', ``edges'', ``edged'', etc)
then as table \ref{tnetrev} shows the second most
popular stem after `network' is `vertic' followed by `edg'. Taking synonyms into account reinforces this picture.  Further, edges
may often be referred to in the  context of the calculation
of some vertex property, such as degree.
\begin{table}[htbp]
\begin{center}
\begin{tabular}{|l|c|c||l|c|c|}
\hline
\textbf{Stem} & {\textbf{Rank}} &  {\textbf{Count}} & \textbf{Stem} &  {\textbf{Rank}} &  {\textbf{Count}} \\ \hline
network & 1 & 254 & number & 11 & 58 \\ \hline
vertic & 2 & 107 & distanc & 12 & 48 \\ \hline
edg & 3 & 86 & model & 13 & 47 \\ \hline
random & 3 & 86 & connect & 14 & 46 \\ \hline
graph & 5 & 81 & data & 15 & 40 \\ \hline
degre & 6 & 78 & link & 16 & 38 \\ \hline
power & 7 & 68 & world & 16 & 38 \\ \hline
lattic & 8 & 67 & hub & 33 & 25 \\ \hline
law & 9 & 65 & point & 38 & 23 \\ \hline
vertex & 10 & 61 & site & 40 & 22 \\ \hline
\end{tabular}
\end{center}
\caption{Table showing the frequencies of the main network
related words in the review of networks \cite{Evans04}. In
calculating the frequencies, `stop words' (such as ``the'')
were removed and then remaining words were stemmed (so the stem
`edg' counts both ``edge'' and ``edges''). The rank is by the
number of occurrences of each word.} \label{tnetrev}
\end{table}

In some cases this focus on vertices is appropriate.  Perhaps,
on the other hand, this predominance of vertex concepts
reflects an inherent bias in the way we humans conceptualise
networks. One way to compensate for our vertex centric view of
the original network is to represent other structures of a
network, here cliques, in terms of the vertices of a new
derived graph. We may then exploit our natural bias in the
analysis of the new derived graph while at the same time
avoiding our propensity for vertices in the original network.

Cliques --– complete subgraphs --- are an important structure
in graph theory. The name originates from representation of
cliques of people in social networks \cite{LP49}.  They have since been used for many purposes
in social networks \cite{LP49,F92,H94,WF94,F96,PS98,EB98,K98,K99,S00,KK02,UCInet02,NMB05,HR05,KH07,B09}. Triads,
cliques of order three, are of particular interest.  One example is the idea that the most important strong ties (in the  language of Granoveter \cite{G73,G83} need to be defined in terms of their membership of triads \cite{F92,K98,K99,KK02,KH07}.

Cliques are also at the centre of some interesting graph
theoretical and algorithmic problems.  Finding the set of all
`maximal' cliques (a clique is maximal only if it is not a
subgraph of another clique) is a good example for which the
Bron-Kerbosch algorithm \cite{BK73} is the classic solution.

Cliques are often used to analyse the general structure of a
network, for example see \cite{SM98,TOA07}.  A particular
application is to use cliques in the search for communities, as in
\cite{PDFV05,KKKS08,YG09a} for example, or equivalently what
are called cohesive groups in social network analysis \cite{WF94,S00,NMB05}.
Alternatively they have been used to produce a model of growing networks
\cite{TO05,TOSHK06}.

Given the importance of cliques we ask if we can shift our focus
away from vertices and onto the cliques of our graph of interest,
$G$, by constructing a ``clique graph'' in which the vertices of
the clique graph represent the cliques of the original graph and
the way they overlap.  Once this has been done, one can use the
standard tools to analyse the properties of the vertices of the
clique graph in order to derive information about the cliques in
the original graph $G$.  There are many such vertex centred tools
but to illustrate the principle we shall look at one complex
example, that of finding communities in networks, the topic of
cohesion in the social networks literature, clustering in the
language of data mining.

The vast majority of community detection algorithms produce a
partition of the set of vertices \cite{F09,POM09}.  That is each
vertex is assigned to one and only one community. These may be
appropriate for many examples, such as those used to illustrate or
test vertex partition algorithms.  However it is an undesirable
constraint for networks made of highly overlapping communities,
with social networks being an obvious case. There one envisages
that the strong ties are formed between friends where there is a
high probability of forming triads through different types of
relationship \cite{KW06,KOSKK07}.  However friendships may be of
different types, family relationships, work collaborations, links
formed through a common sport or hobby.  In this case it makes no
sense to try to assign a single community to each individual but
it does make sense to hypothesise that each triad can be given a
single characterisation, here a single community label. To find
such communities, we will construct the clique graph and then
apply a good vertex partitioning algorithm to the clique graph.
Thus we will illustrate the general central principle of this
paper, namely that a clique graph enables one to avoid the bias of
a vertex centric world to study networks in terms of their cliques
while at the same time exploiting the very same widely available
vertex based analysis techniques to do the analysis at no extra
cost.

In the next section we will look at why it is important to
construct clique graphs with weights and how this may be done.
As an example of how to use vertex based measures on a clique
graph to study cliques in the original graph, in section
\ref{soc} we will construct various overlapping communities.
Finally in section \ref{sdiscussion} we will consider how this
approach can be generalised and it fits in with clique overlap in the literature.

\section{Clique Graphs}\label{scg}

\subsection{Incidence Graph Projections}

\begin{figure}[htb]
\centering
\includegraphics[scale=0.5]{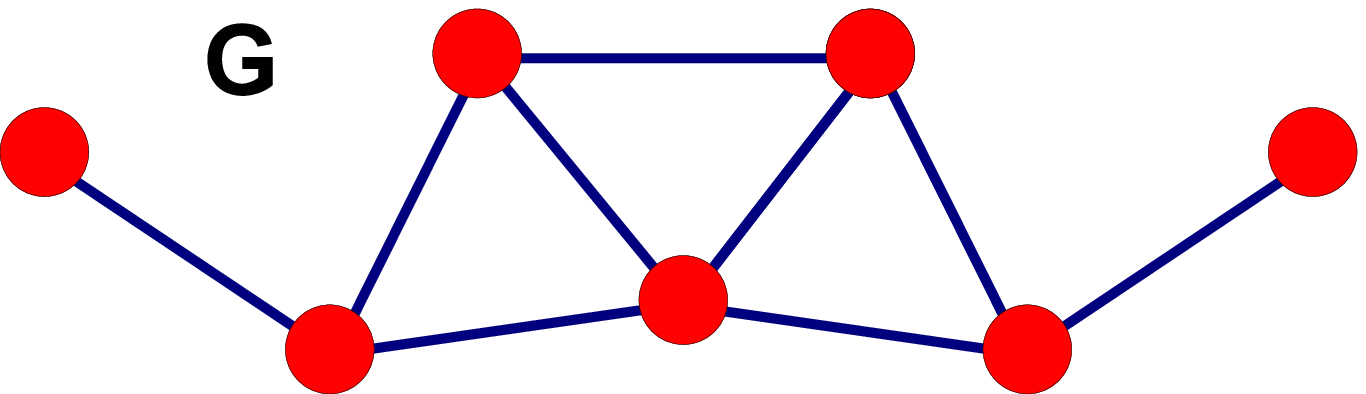}
\\[0.5cm]
\includegraphics[scale=0.5]{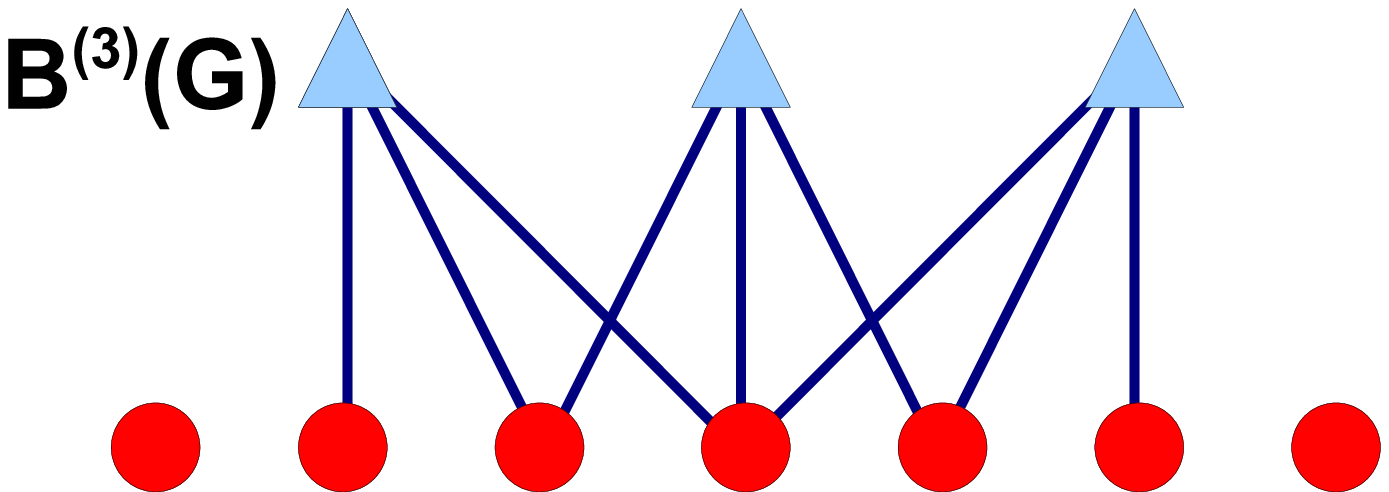}
\hspace{0.05\textwidth}
\raisebox{2mm}{\includegraphics[scale=0.5]{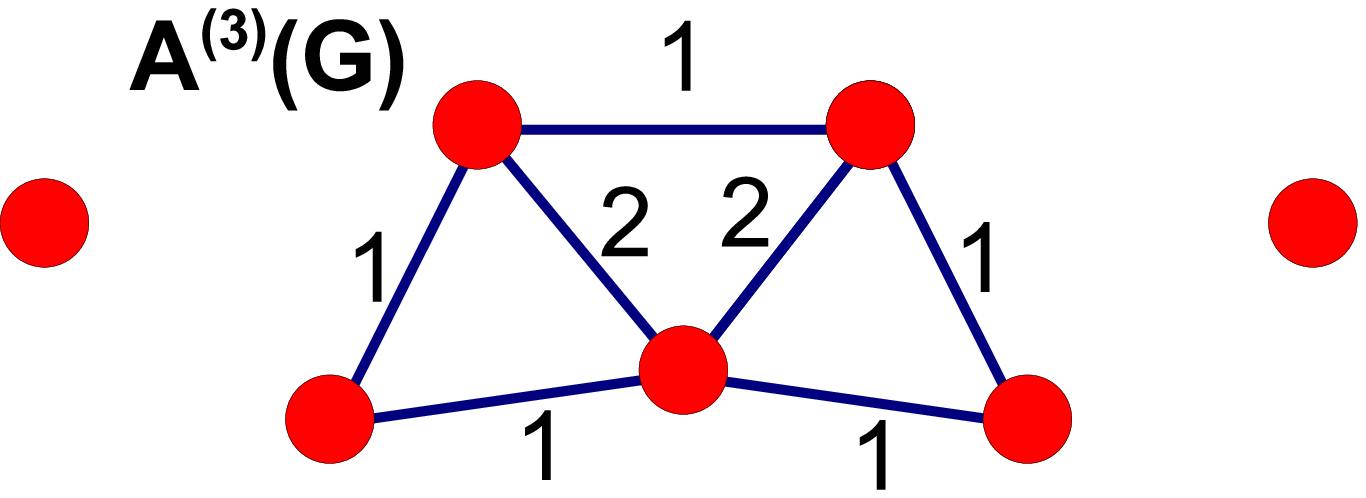}}
\hspace{0.05\textwidth}
\\[0.5cm]
\includegraphics[scale=0.5]{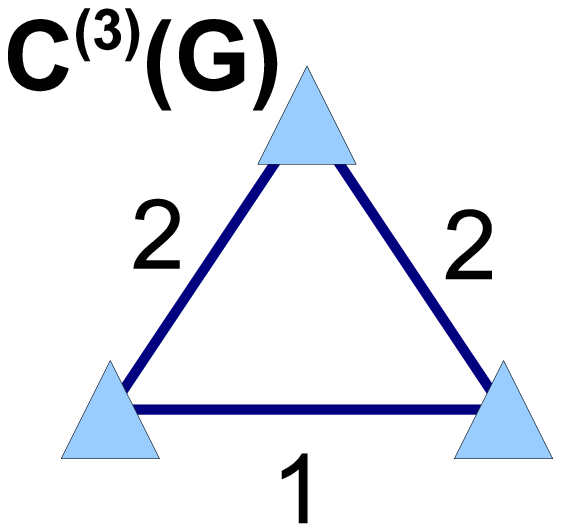}
\hspace{0.05\textwidth}
\raisebox{5mm}{\includegraphics[scale=0.5]{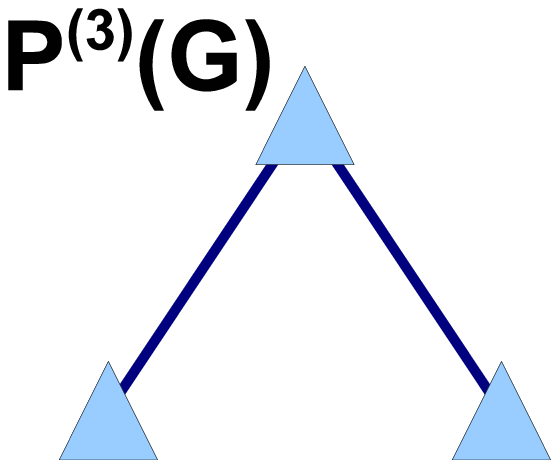}}
\\[0.5cm]
\raisebox{3mm}{\includegraphics[scale=0.5]{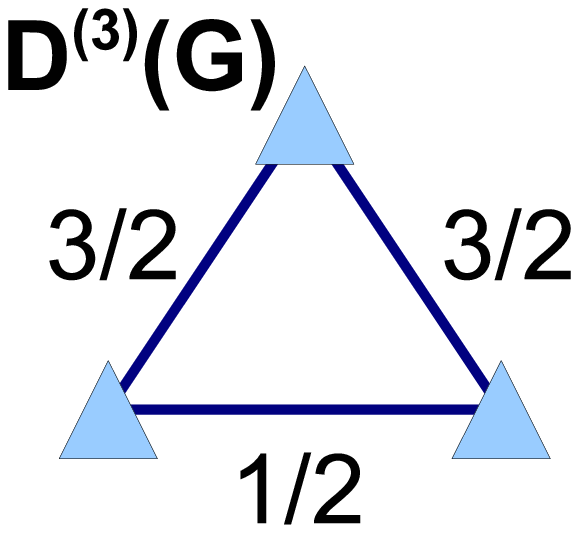}}
\hspace{0.05\textwidth}
\includegraphics[scale=0.5]{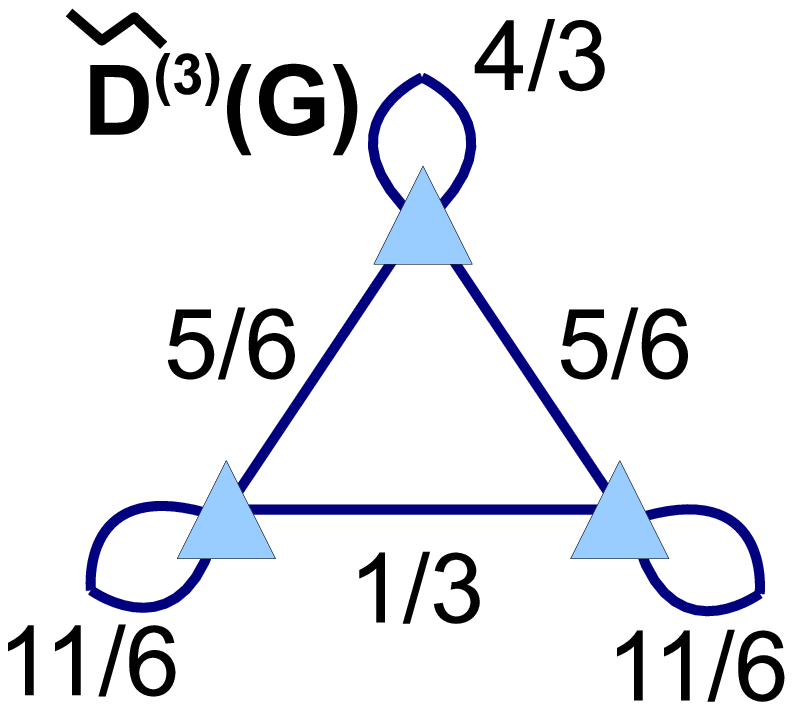}
\caption{An example of the various graphs defined in this paper
for the case of $n=3$. Here the original graph $G$ shown has three
order three cliques, and two vertices in no cliques at all. The
order three clique incidence matrix $\Bthree(G)$ is a bipartite
network whose circle vertices are vertices of $G$ while the three
triangle vertices come from the three order three cliques of $G$.
The incidence matrix can be used to define another graph
$\Athree(G)$ whose unweighted form is isomorphic to a 3-uniform
hypergraph but which is distinct from the original graph $G$. The
clique graphs denoted $\Cthree(G)$, $\Dthree(G)$ and
$\Dtildethree(G)$ correspond to the adjacency matrices defined in
\tseref{adjC}, \tseref{adjD} and \tseref{adjDtilde} respectively.
The unweighted versions of these clique graphs are identical to
the standard line graph of the 3-uniform hypergraph isomorphic to
$\Athree(G)$. The thresholding of the weighted clique graph
$\Cthree(G)$, retaining only edges of weight $(n-1)=2$, produces
$\Pthree(G)$. It is the components of this graph which are used in
the clique percolation method \cite{PDFV05} to define the
communities of $G$.} \label{fC3all}
\end{figure}

Let us consider a simple graph $G$ with vertices drawn from a
vertex set $\Vcal$ and which we will label using mid Latin
characters, $i$, $j$, etc. Now let us consider the set of all
possible order $n$ cliques\footnote{These are distinct from
what are referred to as `$n$-cliques' in the social networks
literature which are not complete subgraphs \cite{WF94,S00}.
However sometimes `$n$-cliques' has also been used to refer to
the order $n$ cliques of interest here, e.g.\ see
\cite{TOA07}.}, $\Ccaln$, for a \emph{single} value of $n\geq2$.
That is $\Ccaln$ is the set of
complete subgraphs of $G$ with $n$ distinct vertices. We will
use early Greek letters, $\alpha$, $\beta$, etc to index these
order $n$ cliques. For instance in the graph $G$ shown in
figure \ref{fC3all} there are three triangles or order three
cliques. For $n=2$ the order two cliques are just the edges of
the original graph $G$.

The relationship between the order $n$ cliques and the vertices of
$G$ can be recorded in an order $n$ clique incidence matrix
$\Bn_{i \alpha}$.  The entries of this $|\Vcal| \times |\Ccaln|$
matrix are equal to $1$ if clique $\alpha \in \Ccaln$ contains
vertex $i \in \Vcal$, otherwise it is $0$
\beq
 \Bn_{i \alpha} = \left\{
 \begin{array}{ccl}
 1 & \mbox{if} & i  \in     \alpha \in \Ccaln \\
 0 & \mbox{if} & i  \not\in \alpha \in \Ccaln
 \end{array}
 \right. \; .
 \label{Bdef}
\eeq
It is useful to define the degree of each vertex $i$ in this
bipartite graph as $k^{(n)}_i$ where
\beq
 \kn_i = \sum_\alpha \Bn_{i \alpha} \, .
 \label{kndef}
\eeq
This is simply the number of order $n$ cliques which contain
vertex $i$ so $k^{(2)}_i$ is
simply the usual definition of the degree of vertex $i$. This order $n$ clique incidence matrix of $G$ may be seen
as the adjacency matrix of a bipartite network, $\Bn(G)$, where
the two types of vertices correspond to the vertices and the order $n$ cliques
of the original graph $G$.  This is shown for the example graph in
figure \ref{fC3all}.

We can construct a new weighted graph $\An(G)$ which is a
subgraph of the original graph $G$  by defining its adjacency
matrix $\Amat$ as follows
\begin{equation}
\An_{ij}
 = \sum_{\alpha\in\Ccaln} \Bn_{i \alpha} \Bn_{j \alpha} (1-\delta_{ij}) \, , \qquad
 \forall \; i,j \in \Vcal \, .
\label{adjA}
\end{equation}
This \tseref{adjA} is the projection of the bipartite incidence
graph $\Bn(G)$ onto a unipartite graph $\An(G)$. The vertex set
of  $\An(G)$ is identical to the original graph $G$. The weight
of edge $(i,j)$ is the number of order $n$ cliques containing
that edge. However, $\An(G)$ is not in general the same as $G$
as any edge not in an order $n$ clique in the original graph
will not be in the $\An(G)$ graph. Each vertex has degree
$\kn_i$ which can be less than the degree of the same vertex
$i$ in the original graph. In particular any vertex of degree
less than $n$ in $G$ will be isolated in $\An(G)$, and any
edges in $G$ incident to such a vertex, will not appear in
$\An(G)$. In the example of figure \ref{fC3all}, the only
difference between $\Athree(G)$ and $G$ are the two edges on
the extreme left and right, neither of which are in any
triangles.

In passing we also note that the
unweighted version of the graph $\An(G)$ is isomorphic to an
$n$-uniform hypergraph \cite{CSBSLHGS07,NO09,NO09a}. Strictly
$\An(G)$ has bipartite relationships between the vertices,
something not explicitly part of a hypergraph definition.
However, our restriction to cliques means that the edges of the
cliques can be deduced from the vertex set of each clique.

More interestingly we could project the bipartite incidence
graph $\Bn(G)$ onto the order $n$ cliques to produce a new
graph $\Cn(G)$.  I will call these ``clique graphs''\footnote{A better term might be
``$n$-regular clique graphs'' or ``order $n$ clique graphs''
since, in the graph theory literature, the overlap between the
set of \emph{all} cliques, not just those of order $n$, are
used to define what are also called clique graphs \cite{H94,MathWorldCG}.
The latter are invariably unweighted whereas weighted edges
will be central in the discussion here.} since each vertex in these new graphs $\Cn(G)$
corresponds to a clique in the original
graph $G$. We will label the vertices of our clique graphs
using the same label $\alpha$ we used for the cliques of $G$.
We could define an edge in a new simple graph $\Ln(G)$ between
vertices $\alpha$ and $\beta$ ($\alpha \neq \beta$) to exist if
there is at least one vertex of the original graph, say $i \in
\Vcal$, which is common to both the order $n$ cliques $\alpha$
and $\beta$. This defines an unweighted simple clique graph,
$\Ln(G)$, which is equivalent to the line graph the $n$-uniform
hypergraph associated with $\An(G)$ \cite{NO09,NO09a}. This
unweighted clique graph $\Ln(G)$ captures the topology of the
clique structure of $G$ but loses a lot of other useful
information.

To retain this information it makes sense to define a weighted
clique graph.  The simplest assignment is to set the weight of
an edge between clique graph vertices $\alpha$ and $\beta$ to
be the number of vertices of $G$ which are common to both
$\alpha$ and $\beta$ order $n$ cliques of $G$.  Thus our first
weighted clique graph, which we will denote as $\Cn(G)$, has an
the adjacency matrix given by
\begin{equation}
C_{\alpha \beta}^{(n)} = \sum_{i} \Bn_{i \alpha} \Bn_{i \beta} (1-
\delta_{\alpha \beta}). \label{adjC}
\end{equation}
Note that we have also chosen to exclude self-loops,
$C_{\alpha\alpha}=0$, as for our order $n$ clique construction the
$\alpha=\beta$ case would always lead to a trivial value of $n$.
The entries $C_{\alpha \beta}$ are therefore an integer between
zero and $(n-1)$ inclusive.  The graph is undirected since
$C_{\alpha \beta}=C_{ \beta\alpha}$.  The $\Cthree(G)$ clique
graph for our example $G$ is shown in figure \ref{fC3all}.

At this point we note that the clique percolation method for
finding communities \cite{PDFV05} may be viewed as counting the
connected components of an unweighted projection of this
weighted clique graph $\Cn(G)$ defined by using a threshold of
$t=(n-1)$ on the weights. That is an unweighted graph
$\Pn(G)$ with adjacency matrix
\begin{equation}
 P_{\alpha \beta}^{(n)} = \left\{ \begin{array}{ccl}
 1 & \mbox{if} & \Cn_{\alpha\beta}\geq (n-1) \\
 0 & \mbox{if} & \Cn_{\alpha\beta}< (n-1)
 \end{array}
 \right.
 \label{adjP}
\end{equation}
So in \cite{PDFV05} only maximal links in $\Cn(G)$ are retained and the communities are then the connected communities of the resulting simple graph.  This seems over restrictive since little of the information in the weights of $\Cn(G)$ has been used yet many methods exist to partition weighted graphs quickly and more effectively.

However this weighted clique graph construction $\Cn(G)$ appears to have a
severe limitation.  Each vertex $i \in \Vcal$ of the original
graph $G$ contributes a total weight of $k^{(n)}_i(k^{(n)}_i-1)/2$
to the edges of $\Cn(G)$. Those which are a member of a large
number of cliques (such as vertices in higher order cliques) will be giving a dominant contribution. If we want
$\Cn(G)$ to be a useful representation of the order $n$ clique
structure of $G$ then it seems much better if we define a clique
graph with different weights on the edges. So we could consider
the following two projections of the incidence matrix onto the
cliques of $G$
\begin{eqnarray}
D_{\alpha \beta}^{(n)} &=& \sum_{i, \kn_i >1} \frac{\Bn_{i \alpha}
\Bn_{i \beta}}{\kn_i -1} (1- \delta_{\alpha \beta}). \label{adjD}
\\
\Dtilde_{\alpha \beta}^{(n)} &=& \sum_{i, \kn_i >0} \frac{\Bn_{i
\alpha} \Bn_{i \beta}}{\kn_i} \label{adjDtilde}
\end{eqnarray}
These adjacency matrices define weighted but undirected clique
graphs, $\Dn(G)$ and $\Dtilden(G)$ respectively with each vertex
$i$ in the original graph $G$ contributing $O(\kn_i)$ to the
weight of these graphs. These weighted line graphs have the
intuitive property that the strength of a vertex $\alpha$ in these
graphs is an integer between 1 and $n$, the order of the cliques
being considered. For $\Dn(G)$ the strength of vertex $\alpha$,
$\str_\alpha = \sum_\beta D_{\alpha \beta}$, is the number of
vertices of $G$ which are in clique $\alpha$ and at least one
other clique.  For $\Dtilden(G)$ the strength is always $n$,
reflecting the fact that each clique has $n$ vertices. These
confirm that we are not giving any one clique too much emphasis.
figure \ref{fC3all} shows the three weighted clique graphs,
$\Cthree(G)$, $\Dthree(G)$ and $\Dtildethree(G)$, for our example
graph.

\subsection{Random Walk Motivation}

There are many other definitions one might try for the weights
of edges in weighted clique graphs, and as with generic
bipartite graph projections, different problems may call for
different definitions \cite{N01a,GL06,ZRMZ07,EL09}.  However
there is another way to motivate the definitions for $\Dn(G)$
and $\Dtilden(G)$ which suggests these are often going to be
the most useful constructions.

Consider an unbiased random walk on the original graph $G$ which
takes place in two stages. First the walker moves from vertex $i$
to any clique $\alpha$ for which the vertex $i$ is a member, that
is $B_{i\alpha}=1$. All cliques attached to $i$ are considered
equally likely in an unbiased walk so this is done with
probability $\Bn_{i \alpha} /\kn_i$. Then the walker moves from
clique $\alpha$ to any vertex $j$ contained in that clique.  Again
all vertices in a clique are considered equally likely so this
step is made with probability proportional to $\Bn_{j \alpha}$.
The process would be identical on the graph $\An(G)$.  It also
corresponds to the natural definition of an unbiased walk on the
$n$-regular hypergraph isomorphic to the unweighted $\An(G)$ in
which walkers move from vertex to hyperedge (the cliques here) to
vertex. Finally it is the natural unbiased walk on the bipartite
incidence graph $\Bn(G)$.  The point about the construction of
$\Dtilden(G)$ is that an unbiased walk on its vertices (which are
the cliques of $G$) preserves the dynamics of the
vertex-clique-vertex walk on the original graph $G$.  Thus any
analysis of the clique graph $\Dtilden(G)$ using a random walk
inspired measure, for example PageRank or modularity optimisation,
will be equivalent to applying these measures to the order $n$
cliques of the original graph without any bias.

By way of comparison, any vertex-vertex random walk done on
$\Cn(g)$ will be equivalent to a biassed vertex-clique-vertex
walk on the original graph $G$ where vertices in many order $n$
cliques (high $\kn_i$) will be preferred by the random walker.

The one unusual point about the walk described for $\Dtilden(G)$
is that it allows processes where walkers can return to the same
point $i \rightarrow \alpha \rightarrow i$ and $\alpha \rightarrow
i \rightarrow \alpha$.  For this two-step process on an undirected
graph it is in some senses natural to allow these.  However,
should one wish to exclude them, as is common in many cases, the
definition of $\Dn(G)$ corresponds to such a process.

Finally, this interpretation suggests that a factor of $1/n$
should be added to \tseref{adjD} and \tseref{adjDtilde} to reflect
the probability of moving from a clique to one of its $n$
vertices. It is an irrelevant constant here but it will be
important if one studies generalisations where cliques of
different orders are considered.


\section{Overlapping Communities}\label{soc}

One can apply any of the many vertex based analysis tools to
clique graphs to get non-trivial information on the cliques in
the original graph.  In this section we look at just one such
example --- the application of vertex partition methods to a
clique graph.

This process assigns a unique community label to each clique in
the original graph. As the first two examples are usually
discussed in terms of vertices, it is natural to associate a
membership function to each vertex.  That is the membership of
a vertex $i$ in a community $c$, say $f_{ic}$, is given by the
fraction of order $n$ cliques containing $i$ which are assigned
to community $c$. That is
\beq
 f_{ic} = \sum_\alpha \frac{B^{(n)}_{i\alpha}}{k^{(n)}_i} F_{\alpha c}
 \label{membfrac}
\eeq
where $F_{\alpha c}$ is the membership fraction for clique
$\alpha$ in community $c$.  Here we have a partition of the set of
$n$ cliques so $F_{\alpha c} = \delta_{c,d}$ if clique $\alpha$ is
assigned to community $d$.  For simplicity, vertices which are not
in any order $n$ cliques, $k^{(n)}_i=0$, are assigned to their own
unique community. Thus vertices may be members of more than one
community and the communities are generally a cover not a simple
partition of the set of vertices. Note that unlike the edge
partition method of \cite{EL09,EL10}, where edges were always assigned
a unique community, here edges also have a natural membership
function but we will not focus on this aspect.

There are many vertex partition methods one can use.  For personal
convenience, the method used in the following examples is the
Louvain algorithm \cite{BGLL08} which gives values for modularity,
$Q(\Amat; \gamma)$, which are close to the maximal value. We use
modify the original form of modularity found in \cite{GN02} and
use \cite{RB06}
\begin{eqnarray}
Q(\Amat; \gamma) = \frac{1}{W}\sum_{C \in \mathcal{P}} \sum_{i,j \in C}
\left[ A_{ij} - \gamma \frac{k_i k_j}{W} \right] \,
\label{modAdef}
\end{eqnarray}
where  $W= \sum_{i,j} A_{ij}$ and $k_i = \sum_{j} A_{ij}$ is the
degree of vertex $i$. The indices $i$ and $j$ run over the $N$
vertices of the graph $G$ whose adjacency matrix is $A_{ij}$. The
index $C$ runs over the communities of the partition
$\mathcal{P}$.  The parameter $\gamma$ may be used to control the
number of communities found \cite{RB06}.

For instance applying the Louvain method to the original karate
club graph with $\gamma=0.3$ usually produces the same binary
split found by Zachary.  With $\gamma=0.5$ the instructor's
faction is split into two with the vertices $\{1,2,3,7,13\}$
assigned to their own community.

One of the big advantages of using modularity is that this may be
interpreted in terms of the behaviour of random walks on the
vertices \cite{DYB08a,LDB08}.  In this language, when we maximise
modularity for the vertices of a clique graph, we can interpret
this as a random walkers on the original graph moving from
vertex to order to vertex and so on. However if we want unbiased
walks on both the original and clique graphs, it is the $\Dn$ and
$\Dtilden$ forms which retain a close relationship.

\subsection{Karate Club}

\begin{figure}
\centering
\includegraphics[width=0.5\textwidth]{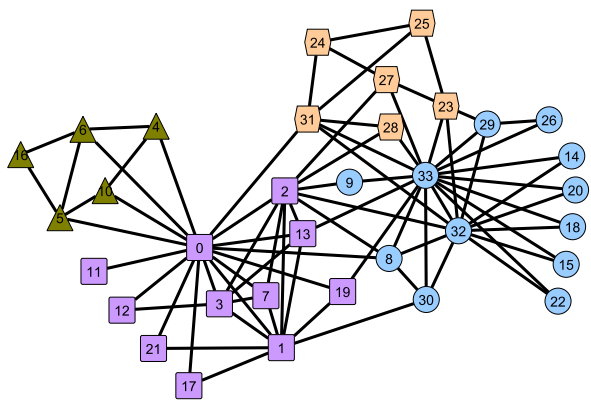}
\caption{Zachary's Karate Club graph \cite{Z77}.  The colour and
shape of vertices indicates the partition of the vertex set which
optimises modularity $Q$ for $\gamma=1.0$ \cite{AK07}. The number
assigned to a vertex is one less than the index used by Zachary so
0 is the chief instructor, 33 the chief officer. The union of
the two subsets on the left (triangles and squares), and the union
of the remaining two subsets (circles and hexagons) form the two
communities found by Zachary \cite{Z77} using the Ford-Fulkerson
binary community algorithm \cite{FF56}.} \label{fkaratevp}
\end{figure}

Zachary \cite{Z77} gave an unweighted, undirected graph of thirty
four vertices, members of a karate club. In this paper, the index
of a vertex is one less than that used by Zachary \cite{Z77}.
Using the Ford-Fulkerson binary community algorithm \cite{FF56}
Zachary split this network into two factions: the instructors
faction centred on vertex 0, and the officers faction, centred on
the vertex numbered 33.  This is shown in figure \ref{fkaratevp}.
Historically the club split into two distinct factions which were
identical to Zachary's artificial partition except for the vertex
numbered 8 here.  This is identical to the actual split in the
Karate club except for vertex 8.

Community algorithms which produce a partition of the vertices
into two sets usually find a split similar to that of Zachary,
suggesting it is an intrinsic feature of the topology of the
network. Subdivisions of these sets to produce three or four
communities are also often found with vertex partition methods,
for example see \cite{AK07,CS10}.

\begin{figure*}
\centering
\includegraphics[width=0.8\textwidth]{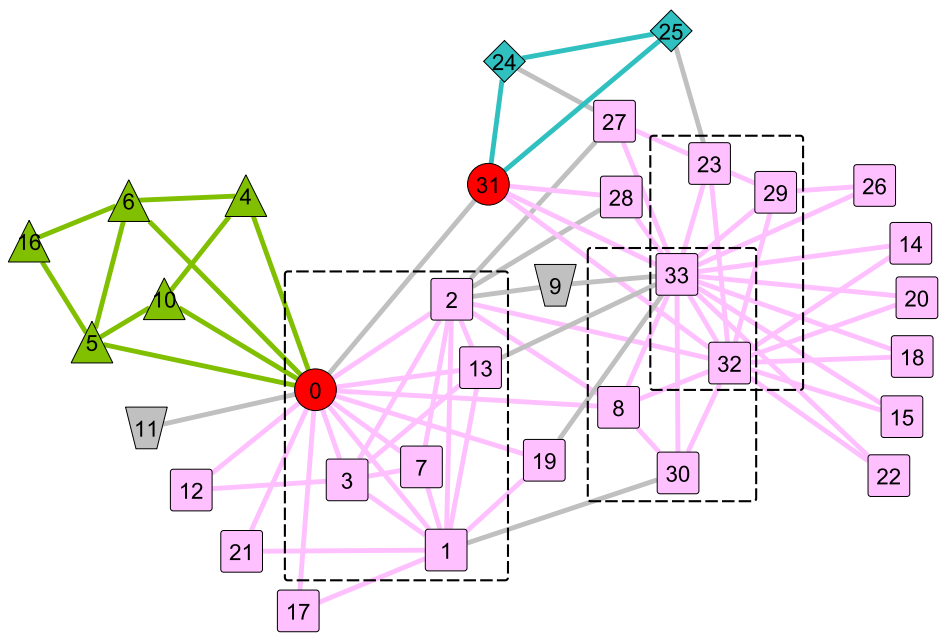}
\caption{The Karate Club of Zachary \cite{Z77}. Vertices 9 and
11 (grey trapeziums) are not in a order three clique, nor are
the eleven edges $(0,11)$, $(0,31)$, $(1,30)$, $(2,9)$,
$(2,27)$, $(2,28)$, $(9,33)$, $(13,33)$, $(19,33)$, $(23,25)$
and $(24,27)$ (grey lines). All other vertices and edges are
part of some three clique.  Two vertices, $0$ and $31$ (shown
as red circles) are the only elements in common between the
three percolating order three clique clusters.  These clusters
are: The three clique $\{24,25,31\}$ (diamond shaped vertices,
blue), the cluster of $\{0,4,5,6,10,16\}$ (triangles and
green), the remaining vertices and edges (square, pink) along
with vertices $0$ and $31$. The rectangular box A contains the
vertices of the two overlapping five cliques (that is
$\{0,1,2,3\}$ plus either $7$ or $13$). Boxes B and C indicate
the two other non-percolating order four cliques,
$\{8,30,32,33\}$ and $\{23,29,32,33\}$.} \label{fkarateperc}
\end{figure*}

Order three cliques play a pivotal role in social network analysis
(see discussion on triads in \cite{WF94} and the examples of
overlapping cliques in \cite{F92,NMB05}), so it seems logical to
consider the case of $n=3$ for the Karate Club. For $n=2$ we would
be constructing the line graphs of the Karate Club which were
considered in \cite{EL09}.  As shown in figure \ref{fkarateperc},
all but two of the thirty four vertices and all but eleven of the
seventy eight edges are in order three cliques. In terms of the
clique percolation protocol of \cite{PDFV05}, the order three
cliques split into three clusters. Equivalently if we remove edges
of weight 1 in the $\Cthree(G)$ graph, we are left with three
components. There is one isolated order three clique,
$\{24,25,31\}$, a second small group involving
$\{0,4,5,6,10,16\}$, and finally one massive community consisting
of all the other vertices plus 0 and 31 again. These three
clusters are connected in terms of all our weighted clique graphs
$\Cthree(G)$, $\Dthree(G)$ and $\Dtildethree(G)$ but they have
just one vertex in common, either 0 or 31.  Thus removing the
weight one edges in $\Cthree(g)$ is equivalent to ignoring this
weak overlap, that is $\Pthree(G)$ has three disconnected
components. Unfortunately this means that the clique percolation
method of \cite{PDFV05} fails to detect the primary binary
division in this graph, one which almost all other methods
successfully detect. Its only success in this context is to
identify the community $\{0,4,5,6,10,16\}$ which is often found if
a community detection method can be set to find more than two
communities.

The higher order cliques of the Karate Club graph are centred
in the two main factions, the two percolating five cliques in
$\{0,1,2,3,7,13\}$ lie in the instructor's cluster, while the
two other two non-percolating order four cliques,
$(8,30,32,33)$ and $(23,29,32,33)$, are entirely within the
officers' cluster. However the simple identification of these
higher order cliques which has achieved the identification of
the core of the two main factions.  The percolation feature of
the algorithm in \cite{PDFV05} adds nothing. Overall, we
conclude that the Karate club graph highlights the weakness of
the clique percolation method \cite{PDFV05}.

However even though the order three cliques are all pervasive
in this example, the basic idea of \cite{PDFV05} and this paper
that cliques can be very informative about community structure
is a good one. One just needs to retain more information than
is done in clique percolation and this is what the weighted
clique graphs achieve.

In terms of our order three clique graphs of the karate club,
applying a vertex partition algorithm to a clique graph assigns to
each vertex a fractional membership of a community, $f_{ic}$ of
\tseref{membfrac}, equal to the fraction of cliques assigned to
that community and which contain the given vertex.

\begin{figure*}
\centering
\includegraphics[width=0.8\textwidth]{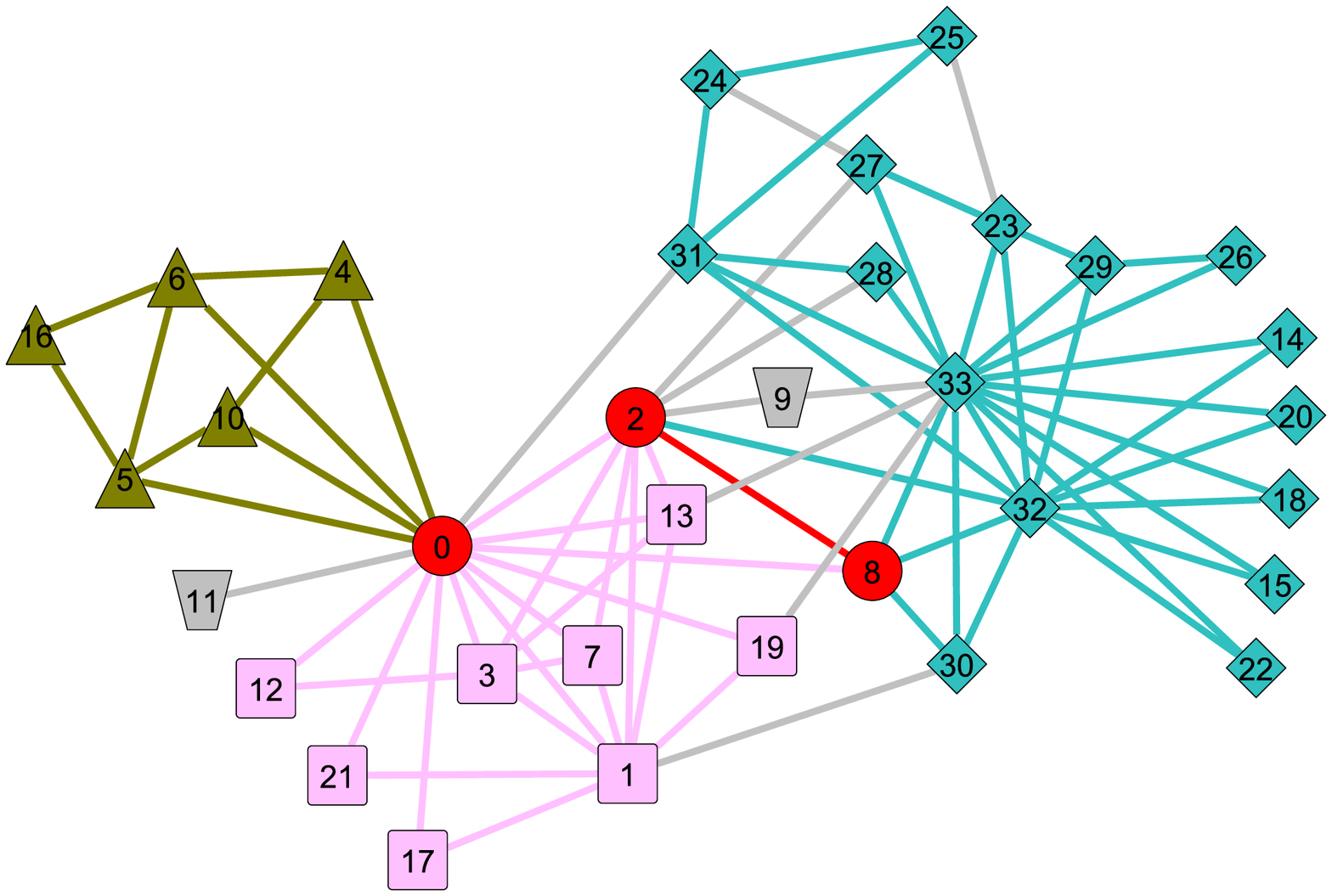}
\caption{The Karate club shown with the partition of the order
three cliques obtained by optimising modularity with $\gamma=0.5$
on the weighted clique graph to $\Dthree(G)$. Three communities of
order three cliques are found. Where a vertex or edge is a member
of order three cliques from only one community, they are given a
unique colour and vertex shape. Vertices 0, 2 and 8 (circles) and
edge (2,8) are members of order three cliques in different
communities and are coloured red. Finally vertices 9 and 11
(trapeziums) and edges $(0,11)$, $(0,31)$, $(1,30)$, $(2,9)$,
$(2,27)$, $(2,28)$, $(9,33)$, $(13,33)$, $(19,33)$, $(23,25)$ and
$(24,27)$  are shown in grey as they are not part of any order
three clique.} \label{fkaratet2c3}
\end{figure*}

The partition of the order three cliques into three communities
found can be interpreted as overlapping communities of vertices
and edges in the original Karate club graph, as shown in figure
\ref{fkaratet2c3}.
For the vertices the community membership is summarised in table
\ref{t2c3h500}.
\begin{table}[hbtp]
\begin{center}
\begin{tabular}{r|l}
 \textbf{Community} & \textbf{Vertices} \\ \hline
 Instructors 1 & 0(78\%), 1, 2(91\%), 3, 7, 8(20\%),  12, 13, 17, 19, 21 \\ \hline
 Instructors 2 & 0(22\%), 4, 5, 6,  10, 16 \\ \hline
 Officers      & 2(9\%), 8(80\%),  14, 15, 18, 20, 22-33
\end{tabular}
\end{center} \label{t2c3h500} \caption{Overlapping
Community structure of the karate club found by partitioning the
vertices of $\Dthree(G)$ using the Louvain method with
$\gamma=0.5$.  If the membership fraction, $f_{ic}$ of
\tref{membfrac} is non-trivial the value is given in brackets
after the index of the vertex.}
\end{table}

The vertices placed in the officers part of the club are placed
completely in one community with the exception of vertices 2
and 8.  Vertex 8 is given only 80\% membership of this faction.
Interestingly though most partitioning methods put this
individual in the officers club, this is the one person which
joined the rival faction in reality.  Though Zachary cites
special circumstances to explain this difference, he also notes
that this person had only a weak affiliation to the officer's
faction. It is therefore not too surprising that our method
does not place this vertex in a unique community. Vertex 2 on
the other hand is assigned only a 9\% membership of the
officers club. This individual was a strong supporter of the
Instructor faction but has significant ties with members of the
Officers faction. Again this does not seem an unreasonable
assignment.

The instructor faction is split into two with vertices 4,5,6,10
and 16 assigned to one community while vertex 0 is given just a
22\% membership of the this group.  Vertex 0 has 78\% of its
order three cliques in the second instructor's faction which
also contains all the remaining vertices with 100\% membership
except for vertex 2 (91\%) and 8 (20\%) as already discussed.

Overall the community structure found by partitioning the
clique graph $\Dthree(G)$ reflects the true nature of the
karate club extremely well.

In the same way we can also study the vertex partitioning of the
clique graph $\Cthree(G)$ for the karate club.  This weighted
clique graph we expected to give too much emphasis to vertices
which are members of many cliques, typically the high degree
vertices. However we find that applying the Louvain method with
$\gamma=0.5$ to partition the vertices of $\Cthree(G)$ that we end
up with two communities. In terms of the original vertices of the
karate club, these are exactly the same as found with $\Dthree(G)$
but where the two instructors communities have been merged.  Thus
though this is still an overlapping community structure, the
overlap (vertices 2 and 8 again) is weak as indicated in table
\ref{t4c3h500}.  So the community structure derived from
$\Cthree(G)$ is also consistent with the binary split of Zachary.
\begin{table}[htbp]
\label{t4c3h500}
\begin{center}
\begin{tabular}{r|l}
\textbf{Community} & \textbf{Vertices} \\ \hline
Instructors  & 0, 1, 2(91\%), 3-7, 8(20\%), 10, 12, 13, 16, 17, 19, 21 \\ \hline
Officers   & 2(9\%), 8(80\%), 14, 15, 18, 20, 22-33
\end{tabular}
\end{center}
\caption{Overlapping Community structure of the karate club
found by partitioning the vertices of $\Cthree(G)$ using the
Louvain method with $\gamma=0.5$). f the membership fraction,
$f_{ic}$ is non-trivial the value is given in brackets after
the index of the vertex.  The binary partition found by Zachary
\cite{Z77} using the Ford-Fulkeson method \cite{FF56} is
identical if we assign vertices 2 and 8 completely to the
community with which they have the largest overlap, the
Instructors and the Officers respectively.}
\end{table}

\subsection{American College Football Network}

Another example that has been used elsewhere \cite{GN02} is the
network formed by teams in a league with each vertex
representing one team with two teams linked if they have played
each other that season. For instance of the 115 teams in
the American College Football Division 1-A in the 2000 season,
all but eight are organised into eleven conferences of various
sizes\footnote{The conference assignment used in \cite{GN02}
appears to be for the 2001 season.  The data used here for the
games played between two teams is based on the file
\texttt{football.gml} downloaded from Newman's website which is
associated with \cite{GN02}.  However the conference
assignments used here have been derived from other sources.
\tsepreprint{See \ref{sfdata} for additional
information.}}. As teams played between 7 and 13 games with an
average of 10.7 games, most teams do not play each other.
However if a team is in a conference then they play the
majority of their games against other teams in the same
conference. For this reason the eleven conferences are readily
apparent as eleven tightly knit subgraphs, each of which
contains cliques of order five or higher making it a useful
test for community detection methods.

\begin{figure*}
\centering
\includegraphics[width=0.8\textwidth]{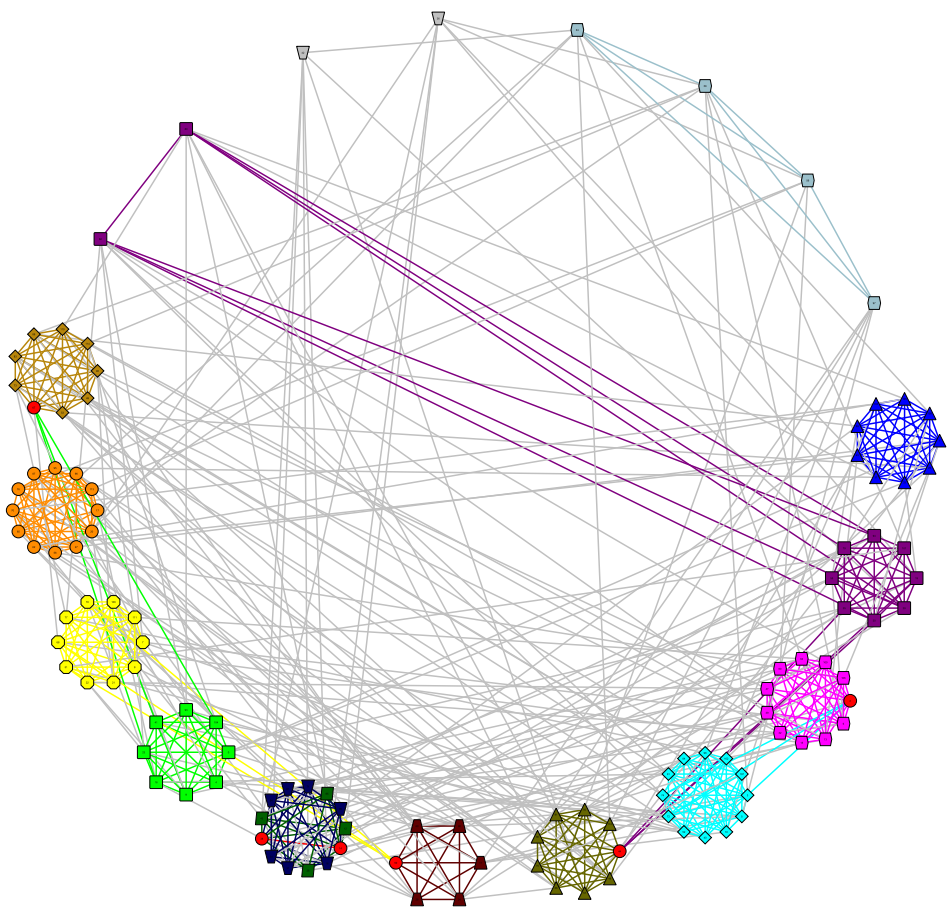}
\caption{Network based on games played in 2000 between teams of
the Division 1-A American College Football league. The community
structure is found using order four clique percolation. Edges and
vertices in a unique community are shown in a colour unique to
that community.  Vertices in the same community are also shown
using the same shape (some shapes are used for two distinct
communities).  Vertices and links not in a order four clique are
shown in grey.  Vertices and edges in more than one community are
shown in red and using circle for the vertices. The teams of each
conference are placed in a small circle which are in turn located
around a large circle (see table \tseref{tconf}). The eight
independents appear as single vertices around the large circle.
The community structure detected by order four clique percolation
matches the conference structure almost perfectly. Note the
conference at about 7 o'clock is split into its two divisions.}
\label{ffootball4perc}
\end{figure*}

The results using order four cliques are very good.  Using
percolation, shown in figure \ref{ffootball4perc}, or vertex
partitioning of either the $\Cfour$ or $\Dfour$ clique graphs
(optimising modularity with $\gamma=1$) gives almost the same
results. This is that each conference corresponds to one, or in
one case, two communities. There is a little overlap, almost all
teams are part of four cliques which involve only teams in their
conference.  The exception is that two independents are deemed
part of the community centred on one of the conferences. A final
community is an isolated clique of four independents. The only
difference between the approaches is that one conference (the
seventh counting clockwise from the conference at 3 o'clock) is
split into its two divisions with percolation and $\Dfour$ but not
with $\Cfour$.

Looking at $5$-cliques vertex partitioning of the $\Cfive$ and
$\Dfive$ clique graphs with $\gamma=1.0$ gives the same structure,
putting all but one team into the correct conference though now
two conferences are split into their divisions. Percolation does
almost as well but one of the conferences is split into three
parts. However as $n$ is raised further, results get rapidly worse
and whole conferences fail to be identified.  This is simply
because these higher order cliques are much rarer in this data
set.

\begin{figure*}
\centering
\includegraphics[width=0.8\textwidth]{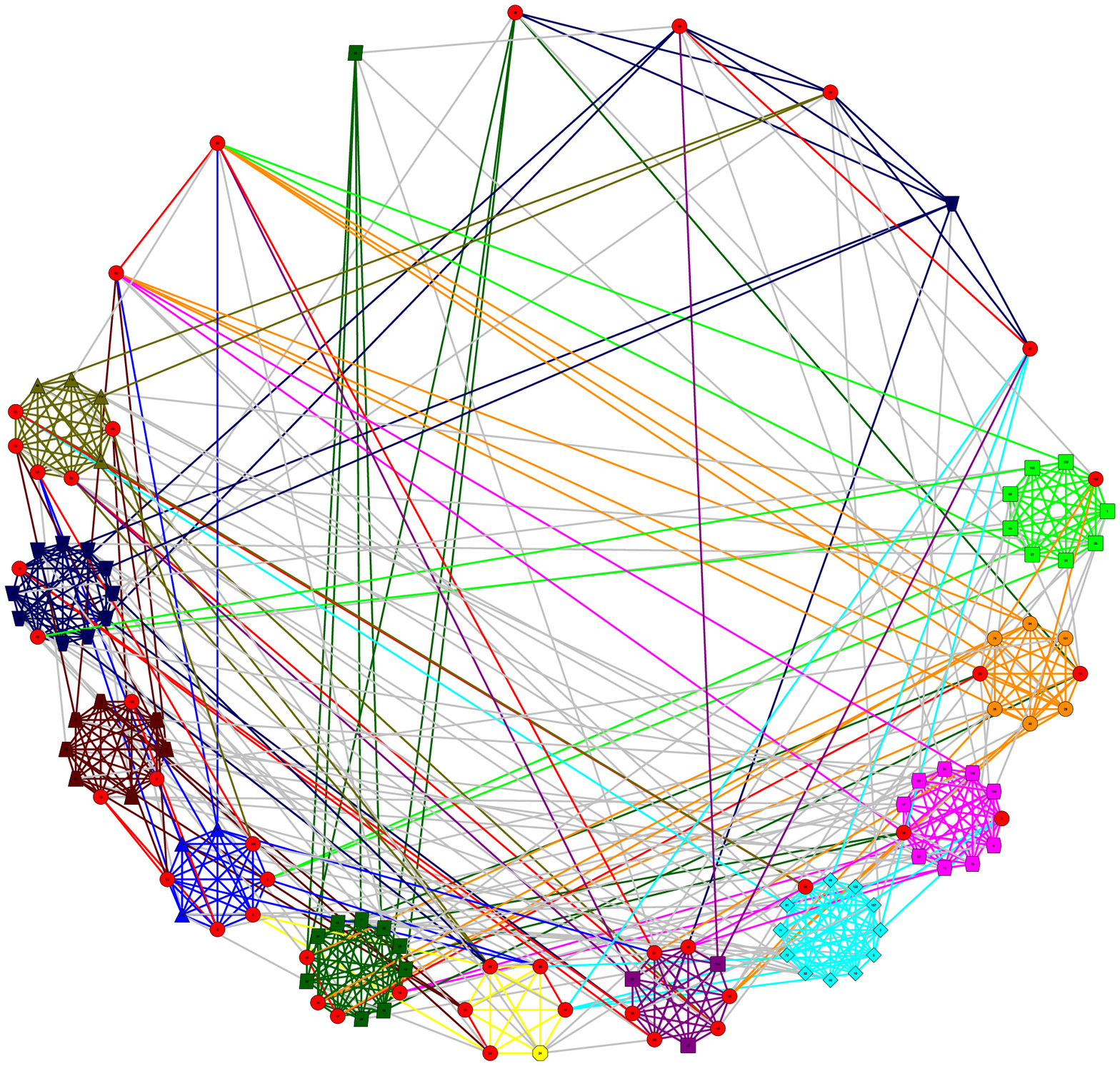}
\caption{Network based on games played in 2000 between teams of
the Division 1-A American College Football league. The vertices
are placed in the same locations as figure \ref{ffootball4perc}.
The community structure detected by vertex partitioning of the
order three clique graph $\Cthree(G)$ clearly identifies the teams
in each conference.  About one-third of conference teams are
members of communities containing teams from other conferences.
However the majority of triangles containing conference teams
contain only other teams in the same conference and so conference
identification is simple.} \label{ffootball3cl1}
\end{figure*}

The real test comes when we consider three cliques for this
American College Football network.  This is a disaster for the
percolation approach as only four communities are identified, two
correspond to one conference each, one is based on the clique of
four independents and all the remaining conference teams are all
in one giant community. However vertex partitioning of both the
$\Cthree$ and $\Dthree$ clique graphs still works, see figure
\ref{ffootball3cl1}.  About two-thirds of the conference teams are
placed in a unique community containing only teams from their
conference and perhaps some independents. For the other third, it
is still true that the vast majority of triangles (at least 79\%)
containing these conference teams contain only other teams in the
same conference.  That is it is easy to classify the overlap as
weak and accurate conference identification remains simple.  Even
the associations seen between some of the independents remain
clear at the 75\% level.

\subsection{Benchmark Graph}

The previous examples have shown that community detection based
on the vertex partitioning of a clique graph can be very
successful, much more than simple clique percolation.  However
all the examples above are well known from the literature which
is dominated by successful methods for finding vertex
partitions of graphs. This means that exemplary networks drawn
from the literature are likely to have an inherent bias towards
those that give `good' results for most vertex partitioning
schemes such as the Louvain method used here.

In fact the situation with these standard examples may be even
more complicated. The definition of a `good' community is
usually taken to be in terms of some reference vertex
partition, Zachary's original split \cite{Z77} into two vertex
sets using the Ford-Fulkerson binary community algorithm
\cite{FF56}, or the association of American College football
teams with their conferences.  A `good' method is defined to be
one which obtains results close to these externally specified
partitions.  Indeed this is what has been done to judge the
clique graph method a success on the previous examples. However
one might argue that a good overlapping community structure
might reveal subtleties missed by simple vertex partitions. For
instance, it is clear that the Instructor in the Karate club
example (vertex 0) is a member of two distinct communities, and
indeed is the only connection between the two. In this sense
the reference partition of the vertices may well not be the
`best' way to describe the community structure in a network.
Unfortunately, it is often not possible to produce a `better'
reference community structure.  Either the data to do this is
not available or there is still a subjective element to any
definition of a better community structure.

Therefore the final example is an artificial benchmark graph
constructed to reflect the overlapping community structures
expected in many situations. Thirty six vertices are placed on
a square grid. Each vertex is visited in turn and two more
vertices are chosen at random, subject to the constraints that
the vertices are distinct, and that all three vertices are
either all in the same row or they are all in the same column.
The three vertices are then connected forming a triangle, using
any existing edge or adding more if needed. Once all thirty six
vertices have been visited we repeat until the desired number
of triangles have been added.  This produces a simple graph.
where every edge and every vertex is part of at least one
triangle.

This benchmark graph can be thought of as a group of thirty six
individuals who work in six different firms and are members of
one of six different social groups (e.g.\ common sports team,
extended family group) outside work.  No two individuals both
work at the same firm and have the same social interests. Of
course this last restraint is somewhat artificial and the
square grid is too simplistic, imposed for purely for
visualisation purposes. Nevertheless it does try to capture the
idea that people are members of more than one community and
their social interactions, here represented by the triangles,
may take place in different communities.  These communities may
not be obvious if one studies just the existence of bilinear
relationships (e.g.\ edges only indicate phone calls were made
or emails were sent) rather than analysing the nature of each
contact.  The aim for a community detection method is to find
the twelve communities, one for every row and one for every
column.

\begin{figure}
\centering
\includegraphics[width=0.8\textwidth]{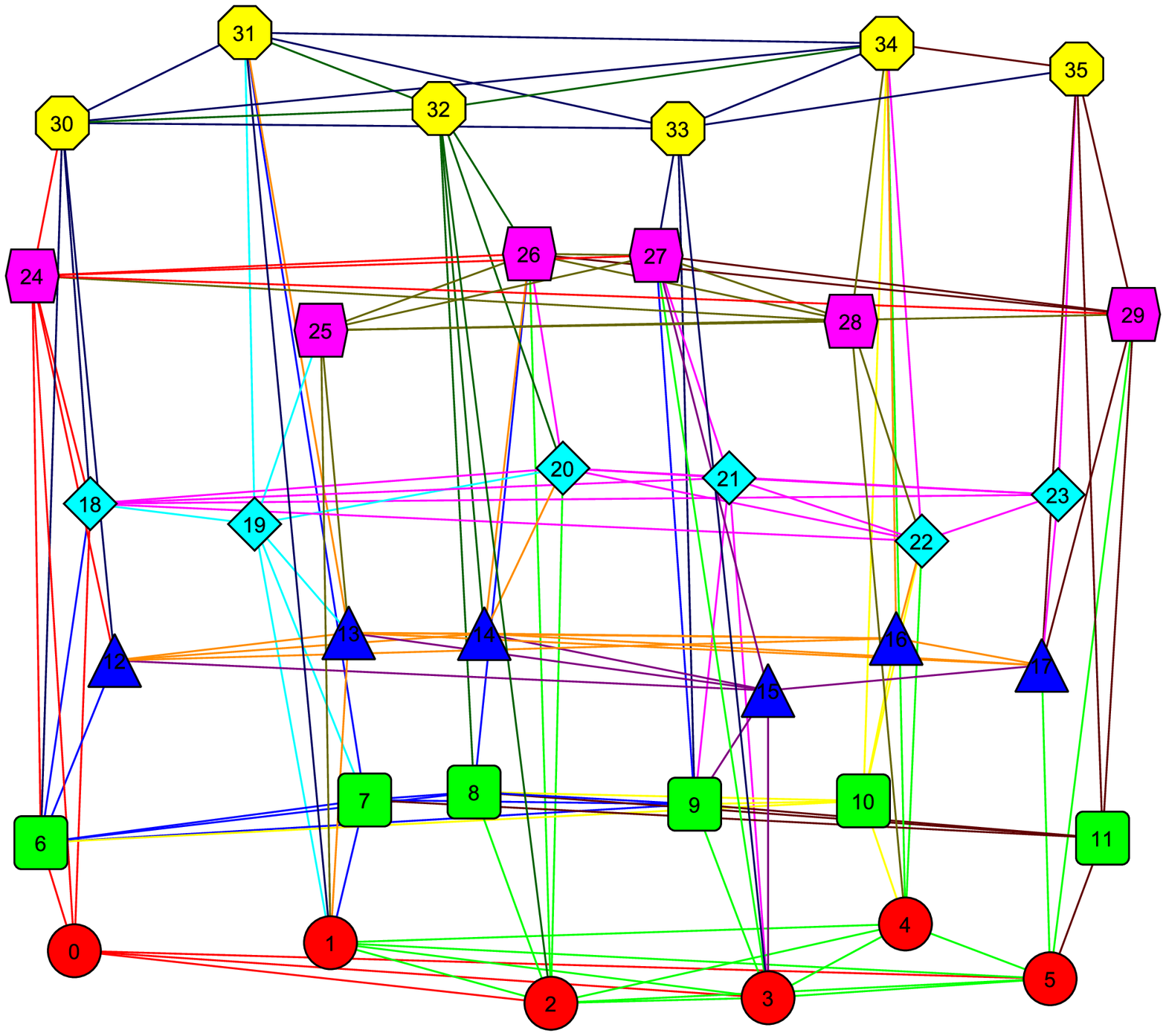}
\caption{A vertex partition and an edge partition of the benchmark graph in which 72 triangles are placed.
The edge partition is found by finding a vertex partition of $\Dtwo(G)$ ---
the weighted line graph $D(G)$ described in \cite{EL09}.
Both partitions are found using the Louvain algorithm to maximise modularity \tseref{modAdef}
with $\gamma=1.2$. On this run, the vertex partition finds the six communities associated with the rows,
indicated by the vertex shapes and colours,
but the row communities are completely missed.
The edge partition finds eleven communities. Most of the lines in each column and most of this n two of the rows
are correctly assigned to a single community, as indicated by the edge colours.}
\label{fTGCGt72vpep}
\end{figure}

Any vertex partition method will fail to find at least half the
structure.  In the example we have used (the Louvain algorithm
applied directly to the vertices) it does seem to be relatively
successful, usually finding 6 or 7 communities, a good
approximation to either the rows or the columns, for $\gamma=0.6$
to $2.0$ in \tseref{modAdef}.  One good example is shown in figure
\ref{fTGCGt72vpep}.

Perhaps surprisingly partitioning the edges by making a
partition of the vertices in the graph $\Dtwo(G)$ (the weighted
line graph $D(G)$ in \cite{EL09}) is not much more successful.
In principle this should also be able to detect the overlapping
communities. The problem here may be that there are also many
rectangles in this artificial benchmark and these be important
when optimising modularity in the weighted line graph.

\begin{figure*}
\centering
\includegraphics[width=0.8\textwidth]{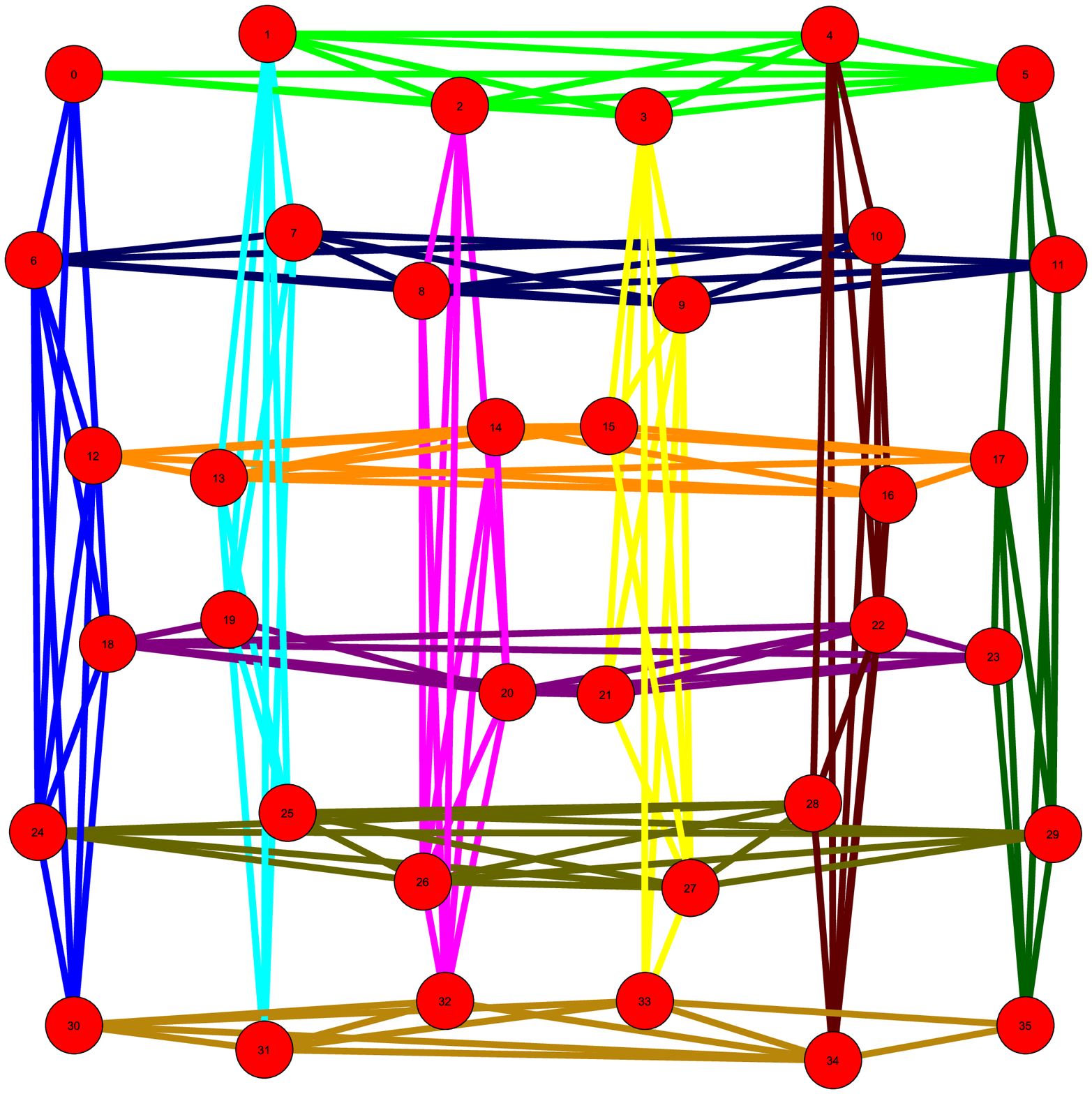}
\caption{The community structure based on the partition of order
three cliques of the same benchmark graph as in figure
\ref{fTGCGt72vpep}.  Produced by applying the Louvain algorithm to
the $\Cthree$ clique graph, maximise modularity with $\gamma=3.0$.
Twelve communities associated with the columns are found matching
the column and row communities perfectly. This is indicated by the
edges in each row having a unique colour, and similar for the
columns.} \label{fTGCGt72t4c3}
\end{figure*}

On the other hand the clique detection method is almost perfect.
Looking at the three cliques, and applying the Louvain method to
both $\Cthree$ and $\Dthree$ clique graphs, both the column and
vertex structure is found almost perfectly as shown in figure
\ref{fTGCGt72t4c3}.  The clique percolation method is also perfect
on this benchmark graph.

\section{Discussion}\label{sdiscussion}

The aim of this paper has been to show that  if one wishes to
focus on the role of cliques in a graph $G$, one may encode
this information as a graph, a clique graph whose vertices
represent the cliques in the original graph  $G$. The advantage
is that there are many well established methods for analysing
the properties of vertices of a graph and for the price of a
simple transformation, these can be applied to obtain the same
information about the cliques.  It avoids the natural bias
towards vertices found in network analysis while exploiting
the same bias by working with clique graphs in order to move
the focus onto the cliques of the original graph $G$.

One of the most important differences between this work and
previous research is that the emphasis here is on the cliques.
Other studies of clique overlap usually retain the focus on the original vertices
and use constructions similar to the $\An_{ij}$ of
\tref{adjA}.  That is the vertices are still the same as the
original graph but now the edge weights carry the information
about clique overlap.  The emphasis here and in \cite{PDFV05}
(and indeed in the clique graphs of \cite{H94,MathWorldCG}) is
to exploit our vertex centric view of graphs and to use a new
graph where the vertices represent the cliques of the original
graph.

The construction of a clique graph is not unique, several
definitions of weighted clique graph are suggested here,
motivated by work on useful projections of bipartite graphs
(for example see \cite{N01a,GL06,ZRMZ07}) and on the case of
order two cliques, the line graphs of \cite{EL09,EL10}.  As
emphasised in \cite{EL09,EL10} the construction of $\Dn$
\tseref{adjD} has the advantage that a random walk on its
vertices retains the dynamical structure of random walks on the
vertices of either the bipartite graph $B$ or the original
graph $G$.

The most obvious limitation so far is that our original graph
$G$ must be simple.  However it is straightforward to define a
second weighted bipartite graph where the entry in the
adjacency matrix $\Bbar_{i\alpha}$ is the weight of the clique
$\alpha$.  There are many ways to define the weight of a clique
based on the weights of the edge, for example see
\cite{SM98,PFPDV07,FAPV07}.  We would consider replacing our
definition of the adjacency matrix of the weighted clique graph
$\Dn$ of \tseref{adjD} by
\begin{eqnarray}
D_{\alpha \beta}^{(n)} &=& \sum_{i, \kn_i >1} \frac{\Bbar^{(n)}_{i \alpha} \Bn_{i
\beta}}{s^{(n)}_i -\Bbar^{(n)}_{i \alpha}} (1- \delta_{\alpha \beta}).
\label{adjDw}
\end{eqnarray}
Here $s^{(n)}_i = \sum_i \Bbar_{i \alpha}$ and $B_{i \alpha}$
is equal to one (is zero) only if $\Bbar_{i\alpha}$ is non-zero
(is zero). This form is again motivated by considering a random
walk that moves from vertex $i$ to clique $\alpha$ to vertex
etc. This approach was used for line graphs $(n=2)$ in
\cite{EL10}.

An important difference between this work and much of the
literature is that I have focused on all cliques of a fixed order
$n$. This can reflect the importance of one particular clique in a
given context.  For instance the triad plays an important role in
social network analysis
\cite{F92,WF94,K98,K99,S00,KK02,NMB05,KH07}. In other
circumstances choosing the order of cliques used may just be a
useful computational freedom, as here and in \cite{PDFV05}.
However it is straightforward to generalise all the constructions
to a situation where the cliques are drawn from a different set of
cliques $\Ccal$, containing cliques of different orders. We just
define a new bipartite incidence matrix $B_{i\alpha}$ which is
one (zero) if vertex $i \in \Vcal$ is in clique $\alpha \in \Ccal$
which is now drawn from some more general set of cliques $\Ccal$.
The clique overlap graph $\An(G)$ defined in \tseref{adjA} is
replaced by
\begin{equation}
A_{ij}(G,\Ccal)
 = \sum_{\alpha\in\Ccal} B_{i \alpha} B_{j \alpha} (1-\delta_{ij}) \, , \qquad
 \forall \; i,j \in \Vcal \, .
\label{adjAgen}
\end{equation}
In fact most work on the overlap of cliques in a graph is based on
$A(G,\Ccalmaxn)$ where $\Ccalmaxn$ is the set of all maximal
cliques whose order is at least $n$, for instance see
\cite{H94,F96,PS98,EB98,UCInet02,HR05}. In principle we could
generalise $\Cn(G)$ $\Dn(G)$ and $\Dtilden(G)$ of
equations \tseref{adjC} \tseref{adjD} and \tseref{adjDtilde} in
the same way, e.g.\ the `co-clique' graph defined in \cite{EB98}
would be $C(G,\Ccalmaxn)$.  However the random walk argument
suggests that the definition of $\Dn(G)$ should now be
\begin{eqnarray}
D_{\alpha \beta}(G,\Ccal) &=& \sum_{i, \kn_i >1}
 \frac{B_{i \alpha}}{(\kn_i -1)} \; \frac{B_{i \beta}}{n_\beta} \; (1- \delta_{\alpha \beta})
 \, , \qquad \alpha,\beta \in \Ccal
 \label{adjDgen}
\end{eqnarray}
where $n_\beta = \sum_i B_{i \beta}$ is the order of the clique $\beta$.

It has been argued that considering only complete subgraphs is too
`stingy' \cite{A73}. So we may be interested in the case where
$\alpha$ is a subgraph of $G$ isomorphic to one of a small set of more
general motifs, subgraphs which are not necessarily regular
graphs. Interesting examples would be those representing cohesion,
such as the $n$-cliques, $n$-clans, $n$-clubs, $k$-plex and
$k$-core structures used in social networks and elsewhere
\cite{WF94,S00,NMB05}.  The incidence matrix $B_{i\alpha}$ may be
defined as before but for this new set of subgraphs and it can be
projected onto vertices or motifs to capture motif overlap. For
instance the generalisation of $A$ and $C$ graphs from a set of
maximal cliques of minimum order as used in \cite{F96}, to
equivalents for a set of motifs was given in \cite{F00}. By the
time we have reached this level of complexity we are essentially
defining hypergraph structures on the set of vertices. On the
other hand, such motif graph constructions are still useful ways
to convey the motif overlap information and, by using a graph to
do so, standard tools may be used to analyse this information.

Such generalisations also suggest how these clique graph
constructions could be adapted for a directed or signed graphs. In
these cases there are many different ways of having connections
between, say, three vertices but we can just keep the relevant
motifs, e.g.\ using the set of triangles regarded as being
balanced in balance theory \cite{WF94,SLT10}.

In all this work we have always considered the overlap of
vertices and motifs.  If the fundamental structure is a graph
$G$ then, in the spirit of \cite{EL09,EL10}, we may want to
define overlap in terms of the edges of $G$. Thus $B_{e\alpha}$
is one if edge $e$ is part of motif $\alpha$. As an example
consider a regular square lattice as the original graph $G$ and
suppose we take a unit square as the motif of interest.  It is
simple to see that the motif graphs formed using the edge
overlap, $\sum_e B_{e\alpha}B_{e\beta}(1-\delta_{\alpha\beta})$
etc, are also square lattices, i.e.\ in term of topology these
edge-motif graphs are just the dual lattice.

Finally we have illustrated one use for clique graphs, that of
detection of overlapping communities, a cover and not a partition
of the original vertices. There has been a recent surge in
interest in this problem, for instance see
\cite{PDFV05,BGM05,LLY06,ZWZ07,NMCM08,G08a,G08b,ABL10,EL09,YG09a,LAV09,LFK09,NO09a,EP07,P09,SSA09,SCG09,WL09,WQWZ09,EL10,SCZ10}.
By way of contrast, the literature on social network analysis,
where clique overlap is better known, is almost entirely focused
on cohesive subgroups which are partitions of the original
vertices, for instance in \cite{Z77,F92,F96,K99,F00,B09,UCInet02}.
This follows in part because of the focus in this area on the graphs which retain
the original vertices such as the $A$'s of \tseref{adjA} and \tseref{adjAgen}.  Since most algorithms produce a partition of the vertices, such as the Johnson Hierarchical Clustering Scheme \cite{J67} (as used in UCInet \cite{UCInet02}), non-overlapping vertex communities are the norm in this area.

The approach suggested here has the advantage that for the price of a simple transformation to produce
a clique graph $C(G,\Ccal)$ or $D(G,\Ccal)$, the much more
extensive work on vertex partition of a graph may be applied to produce an overlapping community structure
without additional work. This may reduce the development time for
a project.  In terms of computational efficiency, the clique
graphs are generally bigger but by how much depends on the
detailed structure of the graph. The speed savings of a good fast
vertex partitioning algorithm, such as \cite{BGLL08,RB08}, may
compensate for the larger size of the clique graph.  The most
important feature though is that this method puts the emphasis on
cliques. It is likely that this approach will be better than other
methods when cliques play a key role. For instance, it was
noticeable that in the benchmark network created out of triangles,
edge partitioning, an alternative overlapping community technique,
was not nearly as effective as the vertex partitioning of clique
graphs.



%
%
%

\section*{Bibliography}

\providecommand{\newblock}{}



\def\tsetrue{T} \def\tsefalse{F} 

\let\tseappendiceson=\tsetrue   

\let\tsewordson=\tsetrue   
\let\tsekarateon=\tsetrue   
\let\tsefootballon=\tsetrue  
\let\tsebenchmarkon=\tsetrue  
\let\tsecolon=\tsefalse  

\if\tseappendiceson\tsetrue

\newpage
\appendix

\if\tsewordson\tsetrue

\section{Alternative Frequency Count}

\begin{table}[htbp]
\begin{center}
\begin{tabular}{c|c|c||c|c|c}
Word & Rank & Count & Word & Rank & Count \\ \hline
network & 8 & 161 & simple & 68 & 33 \\
vertices & 15 & 104 & distribution & 82 & 27 \\
networks & 17 & 90 & scale & 85 & 26 \\
random & 20 & 86 & connected & 91 & 22 \\
degree & 24 & 71 & edge & 106 & 20 \\
graph & 25 & 71 & links & 108 & 20 \\
edges & 28 & 66 & neighbours & 109 & 20 \\
lattice & 29 & 64 & hubs & 121 & 17 \\
power & 30 & 64 & scale-free & 124 & 17 \\
vertex & 34 & 61 & clustering & 129 & 16 \\
distance & 55 & 39 & regular & 155 & 14 \\
small & 64 & 35 & graphs & 207 & 10 \\
world & 66 & 34 & link & 252 & 8 \\
\end{tabular}
\end{center}
\caption{Table showing the frequencies of all the network
related words in the review of networks \cite{Evans04}. In
calculating the frequencies, all words from the text were
included, with no alterations. The rank is by the number of
occurrences of each word, with alphabetical order used to order
words with equal counts.} \label{tnetrevb}
\end{table}

\fi 

\if\tsekarateon\tsetrue
\section{Karate Club}

\subsection{Cliques of the Karate Club}

The index of each vertex is one less than that used by Zachary
\cite{Z77}, so run from 0 to 33.
\begin{itemize}
  \item The highest order of a clique which contains either
      vertex 9 or 11 is two.
  \item All other vertices are in order three cliques.  One
      three clique $\{24,25,31\}$ has no edges in common
      with other order three cliques and only vertex 31 is
      in other order three cliques. The remaining 3 cliques
      split into two groups. Vertices $\{0,4,5,6,10,16\}$
      form a percolating cluster of order three cliques.
      The remaining vertices (all those not mentioned in
      the list so far) along with vertices $0$ and $31$
      form a second percolating cluster. Note this means
      that eleven edges $(0,11)$, $(0,31)$, $(1,30)$,
      $(2,9)$, $(2,27)$, $(2,28)$, $(9,33)$, $(13,33)$,
      $(19,33)$, $(23,25)$ and $(24,27)$  are also not part
      of any order three clique but every other edge is
      part of a triangle.
  \item There are two order four cliques that are not sub
      graphs of cliques of order 5.  These are
      $(8,30,32,33)$ and $(23,29,32,33)$.  Note these are
      not percolating as they share two rather than three
      vertices in common.
  \item There are two order five cliques which are
      percolating around the common four clique of
      $\{0,1,2,3\}$.  Vertices $7$ and $13$ make up the two
      five cliques in this case.
\end{itemize}

\newpage

\subsection{Communities of the Karate Club}

As we see from figure \ref{fkarategnc} the stability in the number
of communities found by the Louvain method suggests that
$\gamma=0.5$ is a reasonable value to study.  In this case we find
three communities of cliques in $\Dthree(G)$ . These correspond
closely to the three communities found for a Louvain partition of
the vertices with $\gamma=0.3$, which in turn is fully consistent
with the binary split of Zachary.
\begin{figure}
\centering
\includegraphics[width=0.8\textwidth]{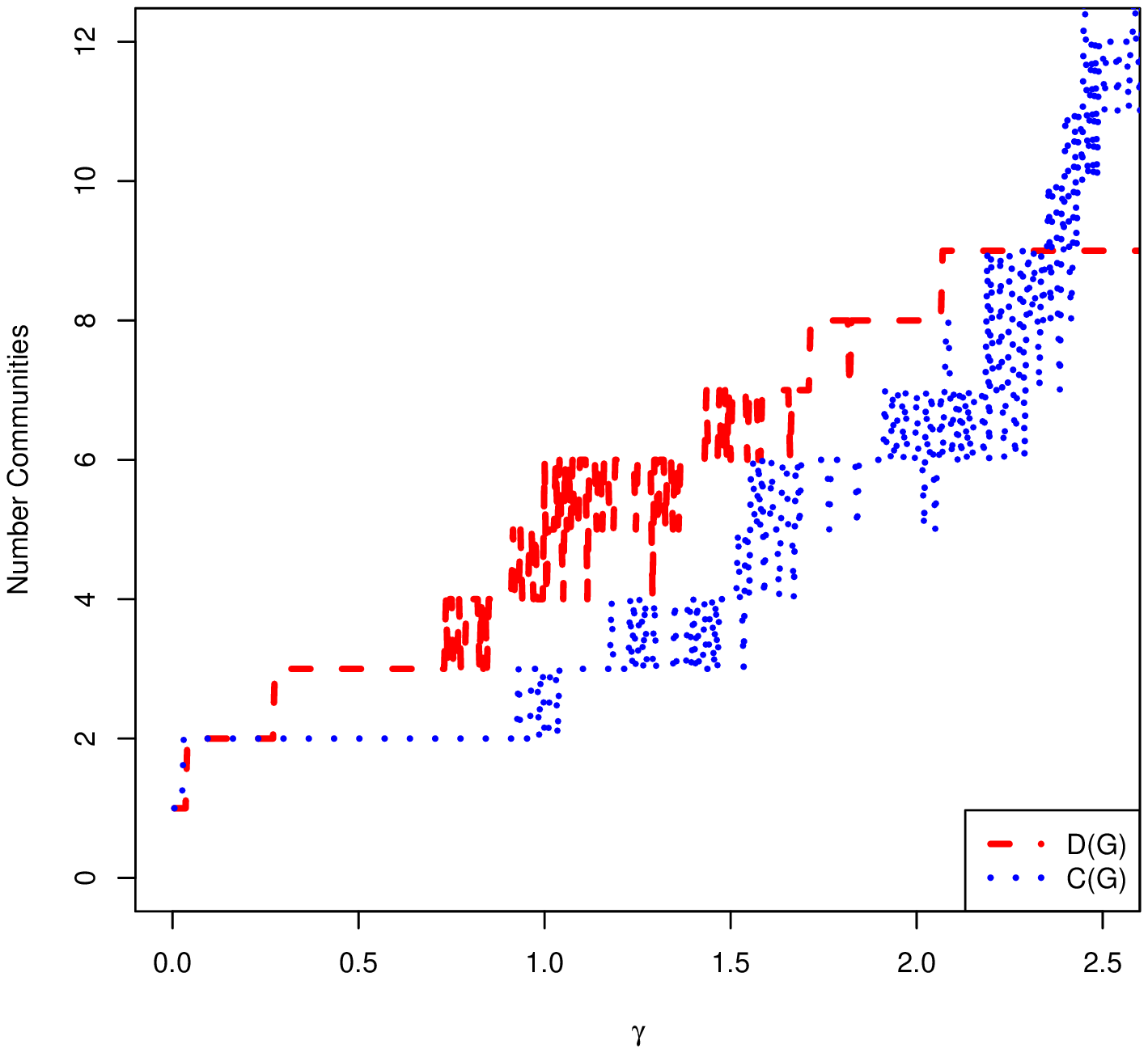}
\caption{The number of communities found maximising the modularity
of \tseref{modAdef} of clique graphs of Zachary's Karate Club
graph as $\gamma$ is varied.  For the clique graphs $\Cthree$
(blue dotted line) and $\Dthree$ (red dashed line).}
\label{fkarategnc}
\end{figure}

The communities referred to in the text are summarised in the
following tables \ref{tZacharyFFapp}, \ref{tBestQapp},
\ref{t4c3h500app} and \ref{t2c3h500app}.

\begin{table}[htbp]
\begin{center}
\begin{tabular}{r|l} \textbf{Community} & \textbf{Vertices} \\
\hline Instructors  & 0-7, 10-13, 16, 17, 19, 21 \\ \hline
Officers     & 8, 9, 14, 15, 18, 20, 22-33
\end{tabular}
\end{center}
\caption{The binary partition found by Zachary \cite{Z77} using
the Ford-Fulkeson method \cite{FF56}. Note that vertex 8 that
is the only one placed in the wrong faction as compared to the
actual split in the club.} \label{tZacharyFFapp}
\end{table}

\begin{table}[htbp]
\begin{center}
\begin{tabular}{r|l}
\textbf{Community} & \textbf{Vertices} \\ \hline
Instructors 1 & 0-3, 7, 11-13, 17, 19, 21 \\ \hline
Instructors 2 & 4, 5, 6, 10, 16 \\ \hline
Officers 1    & 8, 9, 14, 15, 18, 20, 26, 29, 30, 32, 33 \\ \hline
Officers 2    & 23, 24, 25, 27, 28, 31
\end{tabular}
\end{center}
\caption{The vertex partition of the Zachary karate club
\cite{Z77} which produces the largest modularity $Q(A)$
\cite{AK07}.} \label{tBestQapp}
\end{table}

\begin{table}[htbp]
\begin{center}
\begin{tabular}{r|l}
\textbf{Community} & \textbf{Vertices} \\ \hline
Instructors  & 0, 1, 2(91\%), 3-7, 8(20\%), 10, 12, 13, 16, 17, 19, 21 \\ \hline
Officers   & 2(9\%), 8(80\%), 14, 15, 18, 20, 22-33
\end{tabular}
\end{center}
\caption{Overlapping community structure of the karate club found
by partitioning the vertices of $\Cthree(G)$ using the Louvain
method with $\gamma=0.5$. If the membership fraction, $f_{ic}$ of
\tseref{membfrac}, is non-trivial the value is given in brackets
after the index of the vertex.} \label{t4c3h500app}
\end{table}

\begin{table}[htbp]
\begin{center}
\begin{tabular}{r|l}
\textbf{Community} & \textbf{Vertices} \\ \hline
Instructors 1 & 0(78\%), 1, 2(91\%), 3, 7, 8(20\%), 12, 13, 17, 19, 21 \\ \hline
Instructors 2 & 0(22\%), 4, 5, 6, 10, 16  \\ \hline
Officers      & 2(9\%), 8(80\%), 14, 15, 18, 20, 22-33
\end{tabular}
\end{center}
\caption{Overlapping community structure of the karate club found
by partitioning the vertices of $\Dthree(G)$ using the Louvain
method with $\gamma=0.5$. If the membership fraction, $f_{ic}$ of
\tseref{membfrac}, is non-trivial the value is given in brackets
after the index of the vertex.} \label{t2c3h500app}
\end{table}

\newpage
Another example of the community structure found on the karate
club graph, figure \ref{fkaratet4c3app}.
\begin{figure}[htb]
\centering
\includegraphics[width=0.8\textwidth]{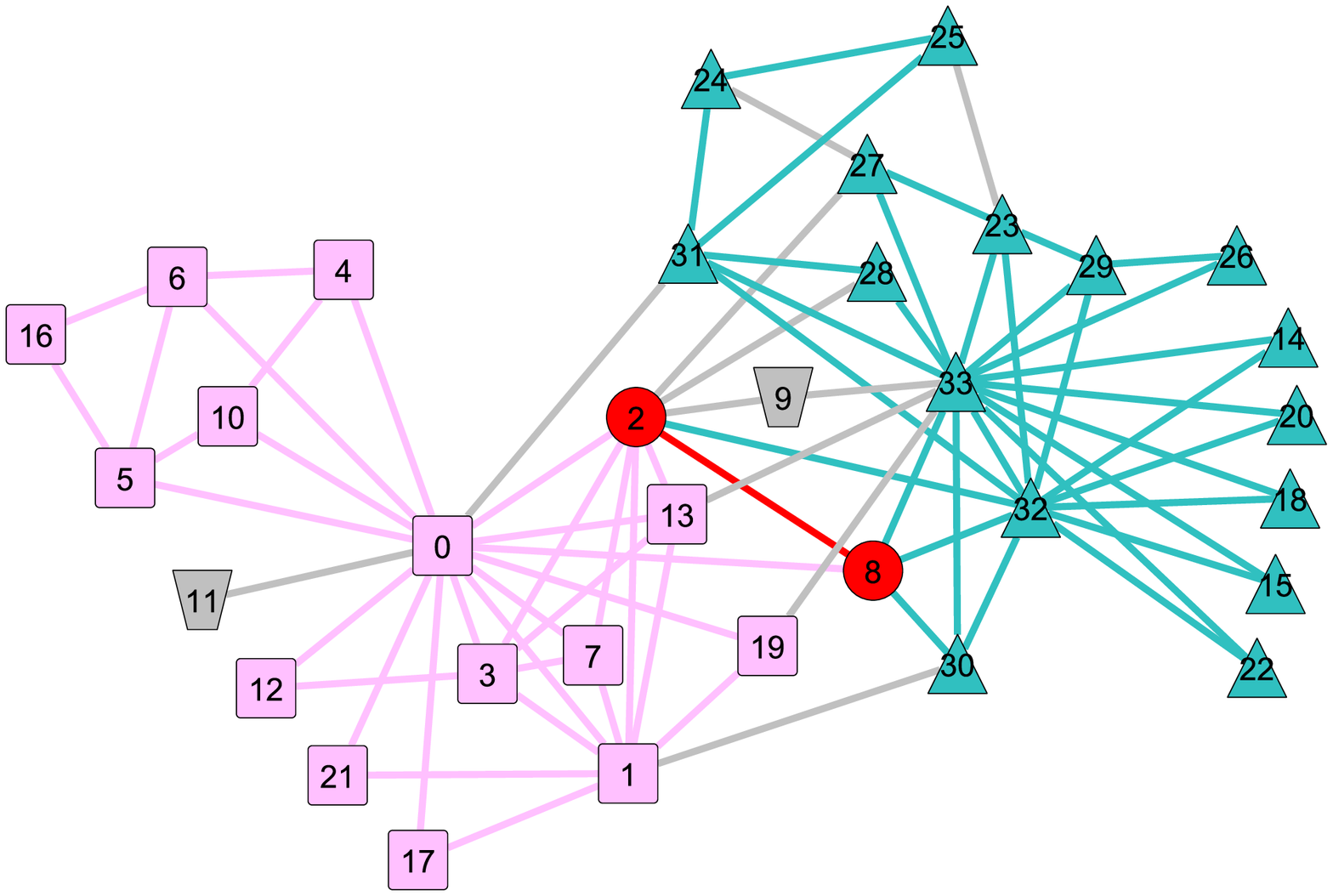}
\caption{The Karate club shown with the partition of the order
three cliques obtained by optimising modularity with $\gamma=0.5$
on the weighted clique graph to $\Cthree(G)$. Two communities of
order three cliques are found.  Where a vertex or edge is a member
of order three cliques from only one community, they are given a
unique colour and vertex shape. Vertices 0, 2 and 8 (circles) and
edge (2,8) are members of order three cliques in different
communities and are coloured red. Finally vertices 9 and 11
(trapeziums) and edges $(0,11)$, $(0,31)$, $(1,30)$, $(2,9)$,
$(2,27)$, $(2,28)$, $(9,33)$, $(13,33)$, $(19,33)$, $(23,25)$ and
$(24,27)$  are shown in grey as they are not part of any order
three clique.} \label{fkaratet4c3app}
\end{figure}

\fi 

\if\tsefootballon\tsetrue
\newpage
\section{American College Football Network}

\subsection{Clique description}

Clique percolation does not work with order three cliques, see
figure \ref{ffootball3cPercapp}. Only four communities are
identified, two correspond to one conference each, one is based on
the clique of four independents and  all the remaining conference
teams are all in one giant community.
\begin{figure}[htb]
\centering
\includegraphics[width=0.8\textwidth]{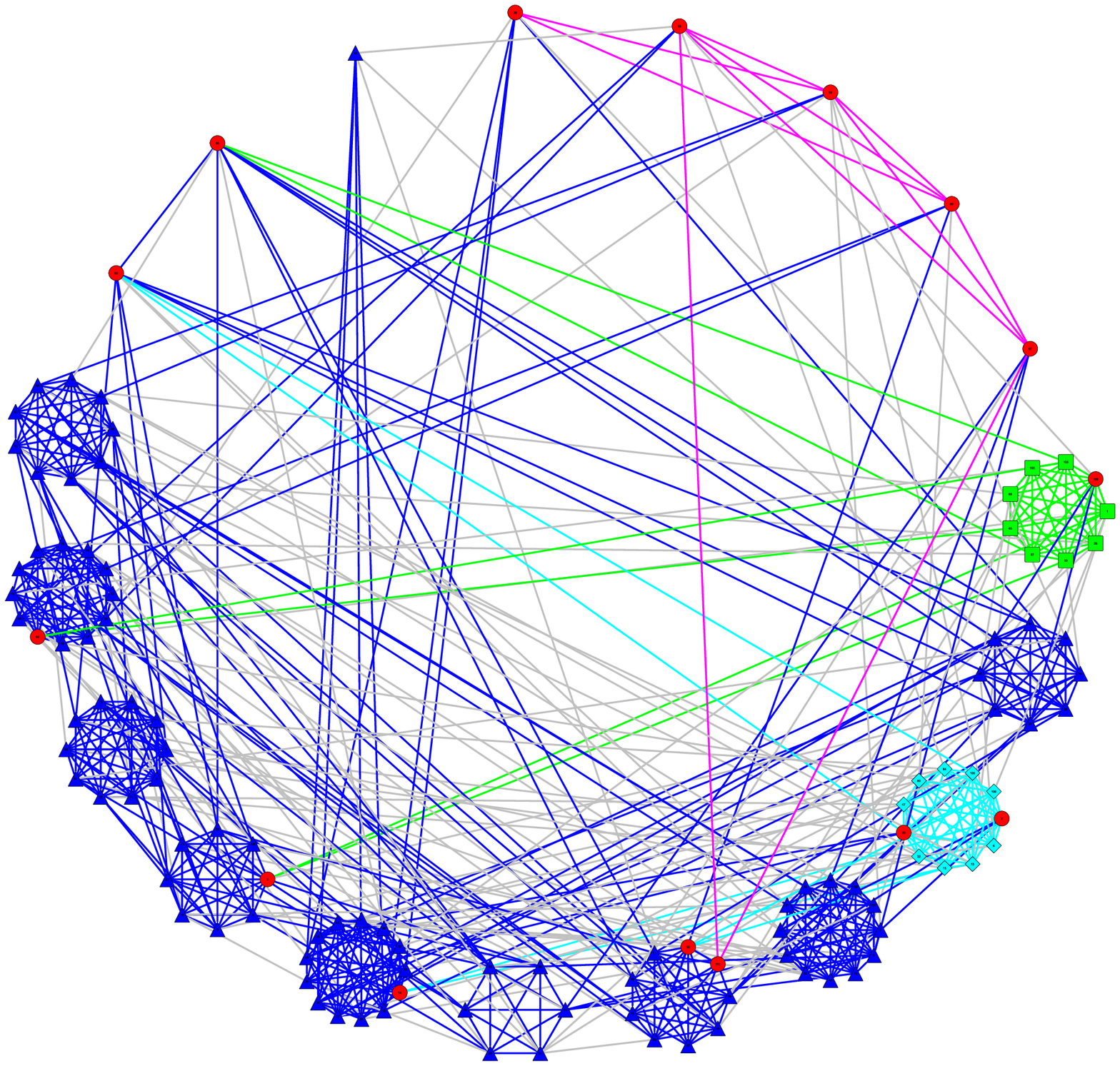}
\caption{Network based on games played between teams of the
Division 1-A American College Football league. The vertices are
placed in the same locations as figure \ref{ffootball4perc}. The
community structure detected by clique percolation of the order
three cliques.  Most of the teams are assigned to one single
community and the conference structure remains largely hidden.}
\label{ffootball3cPercapp}
\end{figure}

\newpage
However vertex partitioning of the $\Cthree$ order three clique
graph still works, see figure \ref{ffootball3cl1app}.  About
two-thirds of the conference teams are placed in a unique
community containing only teams from their conference and perhaps
some independents. For the other third, it is still true that the
vast majority of triangles containing these conference teams
contain only other teams in the same conference.  That is it is
easy to classify the overlap as weak and accurate conference
identification remains simple.
\begin{figure}[htb]
\centering
\includegraphics[width=0.8\textwidth]{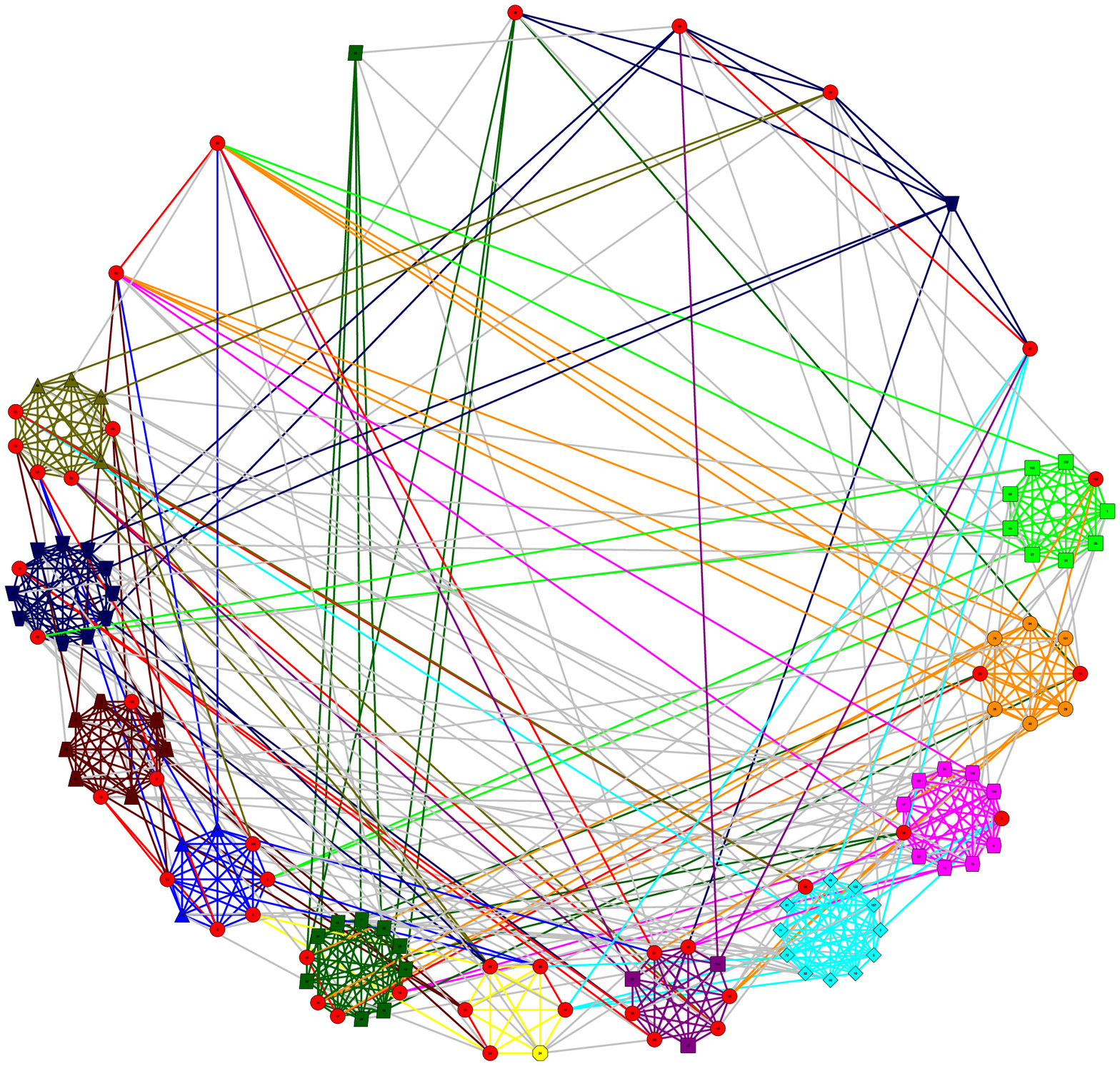}
\caption{Network based on games played between teams of the
Division 1-A American College Football league. The vertices are
placed in the same locations as figure \ref{ffootball4perc}.  The
community structure detected by vertex partitioning of the order
three clique graph $\Cthree(G)$ clearly identifies the teams in
each conference.  About one-third of conference teams are members
of communities containing teams from other conferences. However
the majority of triangles containing conference teams contain only
other teams in the same conference and so conference
identification is simple.} \label{ffootball3cl1app}
\end{figure}

\newpage
Clique percolation is very successful if we work with order four
cliques, see figure \ref{ffootball4perc}. All the conferences are
detected as single groups except for one conference which is split
into two communities. In that case the split reflects the fact
that teams in this conference are also divided into two divisions.
There is a community made from a order four clique of four
independent teams: Middle Tennessee State, Louisiana Monroe,
Louisiana Lafayette and Louisiana Tech.  The first three joined
the Sun Belt Conference the following year while the latter joined
the Western Athletic Conference. Finally a few teams are part of
more than one community. Usually such teams are members of far
many more cliques in their community centred on their conference.
One exception is the case of two independents which are only part
of one community based on one of the conferences: Notre Dame and
Navy are part of the Big East Conference.

\newpage
The vertex partition of the $\Dfour$ order four clique graph
($\gamma=1.0$) finds exactly the same community structure as
clique percolation except it does not split the conferences,
finding one community per conference, see figure
\ref{ffootballt2c4l1app}.
\begin{figure}[htb]
\centering
\includegraphics[width=0.8\textwidth]{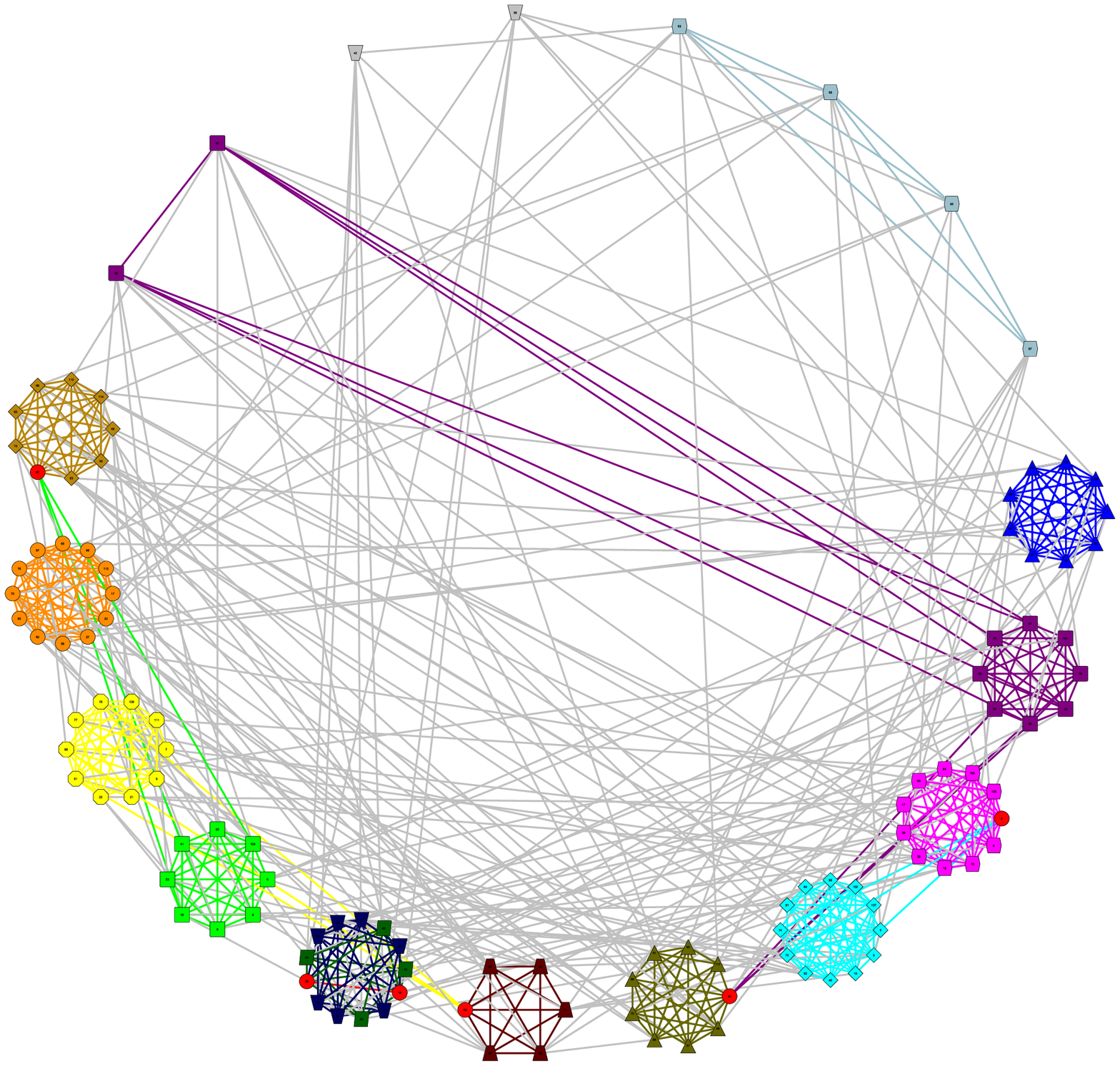}
\caption{Network based on games played between teams of the
Division 1-A American College Football league. The vertices are
placed in the same locations as figure \ref{ffootball4perc}.  The
community structure is determined by optimising modularity for
$\gamma=1.0$ on the $\Dfour$ order four clique graph.  The colours
and shapes are chosen in the same manner as figure
\ref{ffootball4perc}. The community structure detected using this
vertex partition of the order four clique graph $\Dfour$ is
identical to that found using order four clique percolation.  It
matches the conference structure almost perfectly.}
\label{ffootballt2c4l1app}
\end{figure}

\newpage
The vertex partition of the $\Cfour$ order four clique graph
($\gamma=1.0$) also a good match to the conference structure, see
figure \ref{ffootballt4c4l1app}.
\begin{figure}[htb]
\centering
\includegraphics[width=0.8\textwidth]{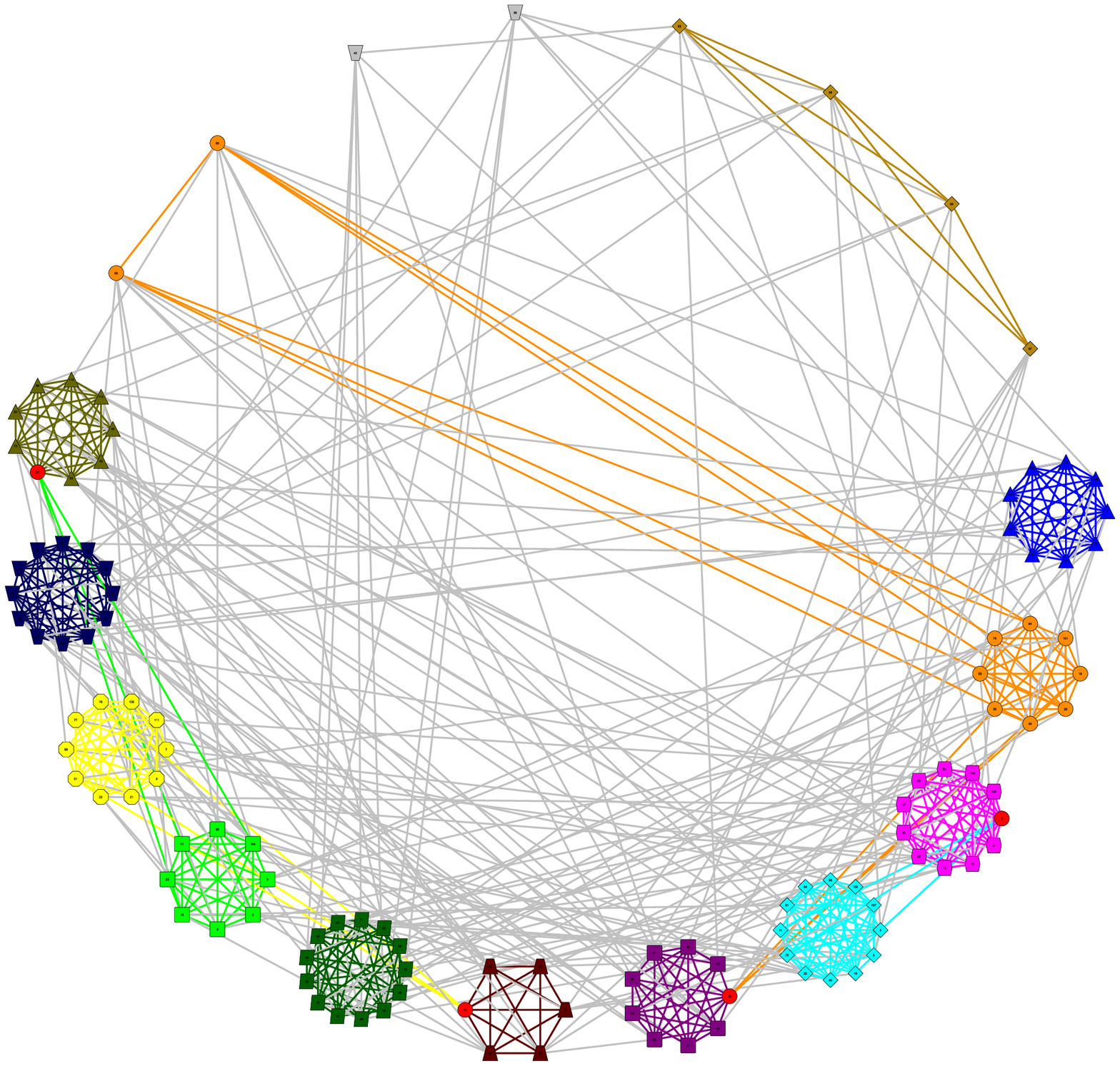}
\caption{Network based on games played between teams of the
Division 1-A American College Football league. The vertices are
placed in the same locations as figure \ref{ffootball4perc}.  The
community structure is determined by optimising modularity for
$\gamma=1.0$ on the $\Cfour$ order four clique graph.  The colours
and shapes are chosen in the same manner as figure
\ref{ffootball4perc}. The community structure detected using this
vertex partition of the order four clique graph $\Dfour$ is almost
identical to that found using order four clique percolation. The
only difference is that the seventh conference (counting clockwise
from the first), the Mid-American, is not split into two
divisions. The community structure detected using this vertex
partition of the order four clique graph $\Dfour$ matches the
conference structure almost perfectly.} \label{ffootballt4c4l1app}
\end{figure}

For order five clique percolation, we find 15 communities, each
of which is entirely within one of the 11 conferences though
one conference team, Alabama Birmingham in Conference USA, is
now no longer part of a order five clique. No independent is in
a order five clique. The method correctly splits two of the
conferences into their two divisions but finds a single
community for two other conferences with divisions.  The one
flaw is that it splits one other conference, which has two
divisions, into three distinct communities. By way of
comparison the vertex partition of the $\Cfive$ order five
clique graph ($\gamma=1.0$) gets exactly the same results
except it does not split one conference into three communities.

\newpage

For order $n$ cliques as we raise  $n$ from 6 upwards, too few
teams are in cliques and an increasing number of conferences are
not identified, see for example figure \ref{ffootball6Percapp}.

\begin{figure}[htb]
\centering
\includegraphics[width=0.8\textwidth]{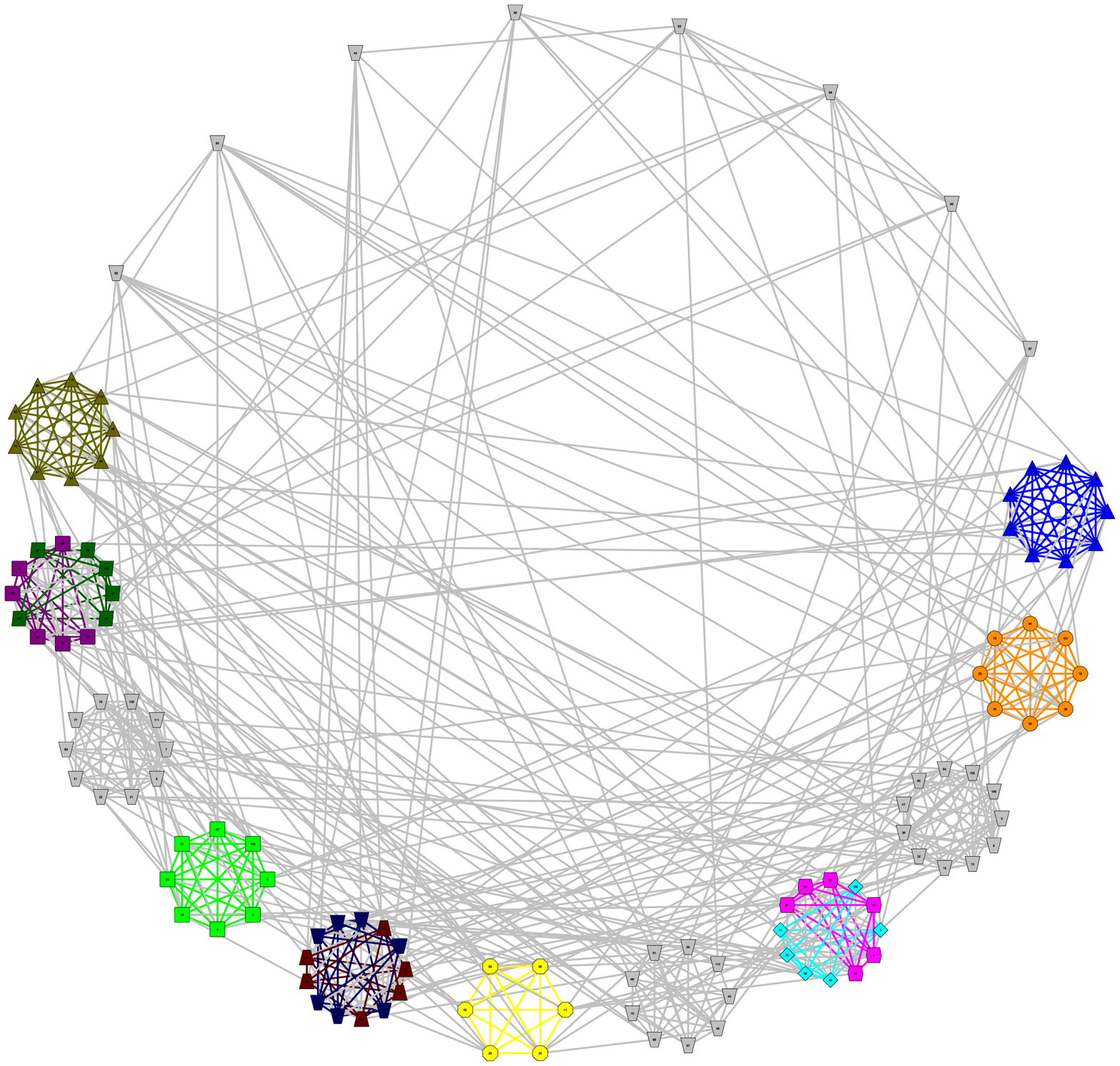}
\caption{Network based on games played between teams of the
Division 1-A American College Football league. The vertices are
placed in the same locations as figure \ref{ffootball4perc}.  The
community structure is determined using 6-clique percolation. The
colours and shapes are chosen in the same manner as figure
\ref{ffootball4perc}. Now the three of the conferences are no
longer detected though the substructure in three other conferences
are detected.} \label{ffootball6Percapp}
\end{figure}

\newpage
\subsection{Data Sources}\label{sfdata}

The links between teams are derived from the file
\texttt{football.gml} downloaded from Newman's web site
(\verb$http://www-personal.umich.edu/~mejn/netdata/$).  This was
compiled by Girvan and Newman \cite{GN02} and gives a link for
every Division I-A game of the American College Football league
during Fall 2000. There is an edge for every game has been played
between two teams.

However, there are two issues with the original file. First three
teams met twice in one season, typically because of conference
finals.  The pairs were: teams 3 (Kansas State) and 84 (Oklahoma),
teams 99 (Marshall) and 14 (Western Michigan), and
teams 27 (Florida) and 17 (Auburn).  For this work only one edge has been created for
these three pairs and the graph is simple.

Secondly, the assignments to made to conferences (the
\texttt{value} tag in the original file) seems to be for the
2001 season and not the 2000 season. In particular the Big West
conference existed for football till 2000 while the Sun Belt
conference was only started in 2001. The new conference
assignments used in this paper are given in table \ref{tconf}.
\begin{table}[htb]
\begin{center}
\begin{tabular}{c|l||c|l}
 Index & Conference & Index & Independent \\ \hline
  0 & Atlantic Coast & 11 & Notre Dame \\
  1 & Big East & 12 & Navy \\
  2 & Big Ten & 13 & Connecticut     \\
  3 & Big Twelve & 14 & Central Florida \\
  4 & Conference USA & 15 & Middle Tennessee State \\
  5 & Big West &  16 & Louisiana Tech     \\
  6 & Mid-American & 17 & Louisiana Monroe    \\
  7 & Mountain West & 18 & Louisiana Lafayette   \\
  8 & Pacific Ten \\
  9 & Southeastern \\
 10 & Western Athletic \\
 \end{tabular}
 \end{center}
\caption{Table listing the Division 1-A American College Football
conferences for the Fall 2000 season.  The indices refer to those
used in this paper. In the figures showing the network based on
games played in the 2000 season, teams from each conference are
placed in order clockwise in a large circle with conference 0
(Atlantic Coast) at 3 o'clock and the last conference (10, Western
Athletic) just after 9 o'clock.}
 \label{tconf}
 \end{table}

\newpage
Using the conference assignments given in table \ref{tconf} for the
2000 season means that corrections were made to the values given in the file
\texttt{football.gml} downloaded from Newman's web site
for Western Athletic and Independent teams.  In addition the following changes were needed:-
\begin{itemize}
\item N.Texas (v11) is in conf 5 Big West (not Sun Belt, GN
    conference 10)

\item Arkansas State (v24) is in conf 5 Big West (not Sun
    Belt GN conference 10)

\item Boise State (v28) is in conf 5 Big West (not Western
    Athletic GN conference 11)

\item Idaho (v50) is in conf 5 Big West (not Sun Belt, GN
    conference 10)

\item Louisiana Tech (v58) is in conf 16, an Independent
    (not Western Athletic GN conference 11)

\item Louisiana Monroe (v59) is in conf 17, an Independent
    (not Sun Belt GN conference 10)

\item Middle Tennessee State (v63) is in conf 15, an
    Independent (not Sun Belt GN conference 10)

\item New Mexico State (v69) is in conf 5 Big West (not Sun
    Belt, GN conference 10)

\item Utah State (v90) is in conf 5 Big West (not
    Independents, GN conference 5)

\item Louisiana Lafayette (v97) is in conf 18, an
    Independent (not Sun Belt GN conference 10)

\item Texas Christian (v110) is in Western Athletic conf 10
    (not Conference USA GN conference 4)

\end{itemize}
A sample of the games have been checked using the results given
on the ``College Football Data Warehouse''
(\texttt{www.cfbdatawarehouse.com}) and using various Wikipedia
entries for the different conferences and teams.

\fi 

\if\tsebenchmarkon\tsetrue
\newpage
\section{Further Benchmark example}

Below are some more examples of communities found on the
benchmark graph.

\begin{figure}[htb]
\centering
\includegraphics[width=0.8\textwidth]{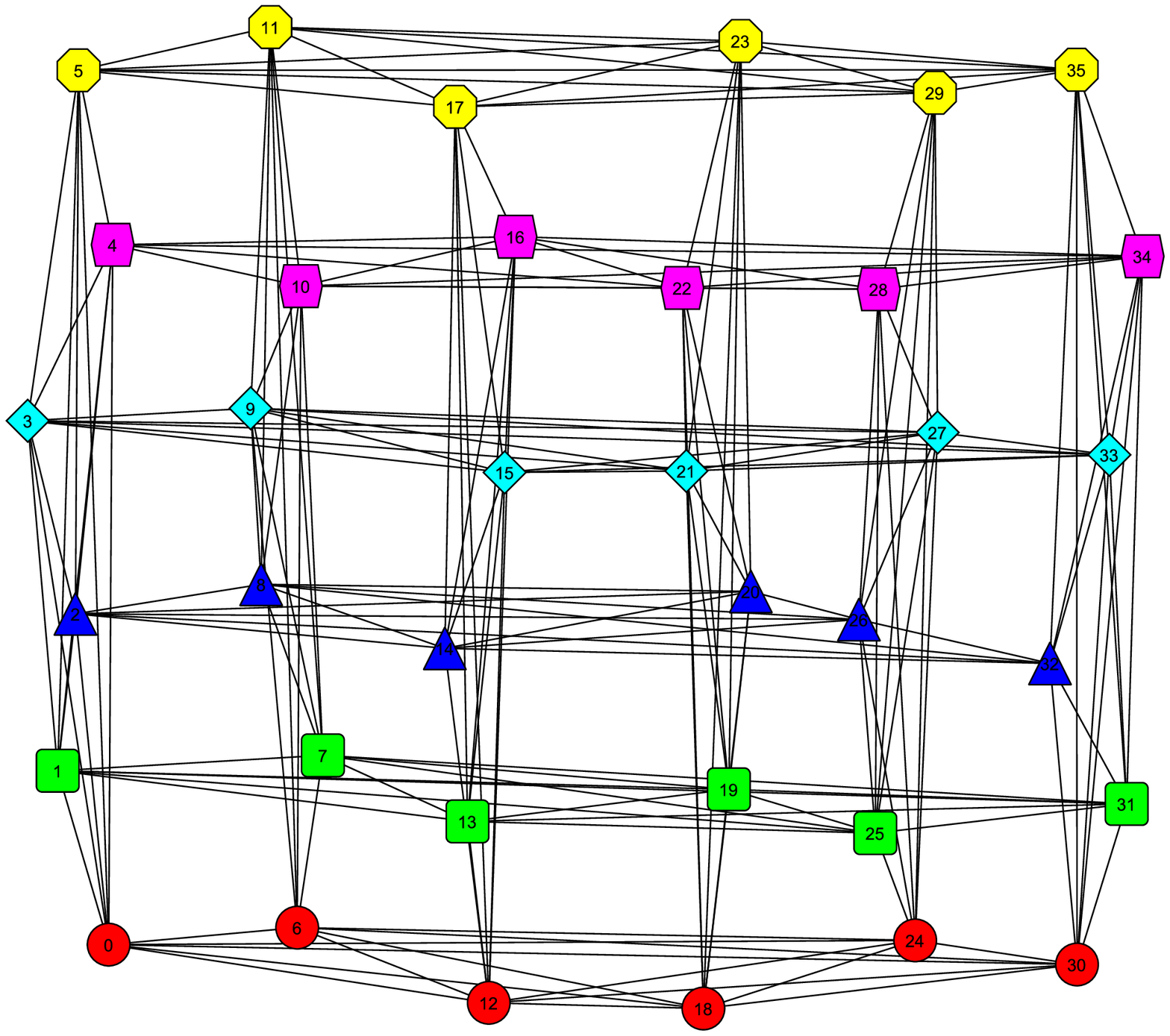}
\caption{A vertex partition of the benchmark graph with 216 triangles using the
Louvain algorithm to maximise modularity with $\gamma=1.0$. On
this run, the six communities associated with the columns are
found, indicated by the vertex shapes and colours,
but the row communities are completely missed.}
\label{fTGCGt216vp}
\end{figure}

\newpage
The community structure based on the partition of order
three cliques of the same benchmark graph as in figure
\ref{fTGCGt216vp}. Produced by applying the Louvain algorithm to
the $\Dthree$ clique graph, maximise modularity with $\gamma=2.5$.
Twelve communities were produced on this run, matching the column
and row communities perfectly. This is indicated by the edges in
each row having a unique colour, and similar for the columns. The
one flaw is that the edges in the third column are assigned to two
communities, one unique to that column, and the other used for the
third row from the top.
\begin{figure*}
\centering
\includegraphics[width=0.8\textwidth]{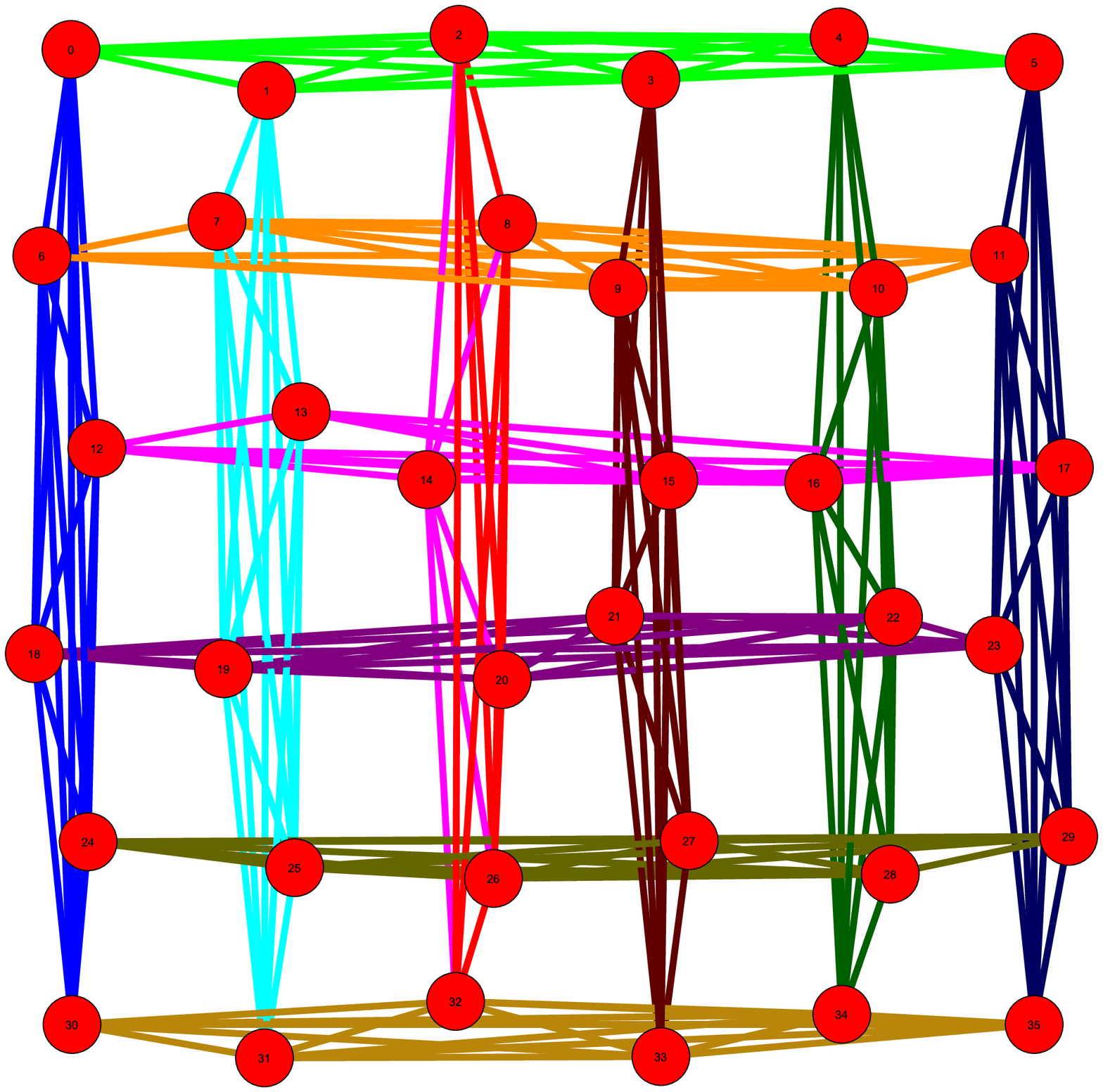}
\caption{The community structure based on the partition of order
three cliques of the same benchmark graph as in figure
\ref{fTGCGt216vp}. Produced by applying the Louvain algorithm to
the $\Dthree$ clique graph, maximise modularity with $\gamma=2.5$.
Twelve communities were produced on this run, matching the column
and row communities perfectly. This is indicated by the edges in
each row having a unique colour, and similar for the columns. The
one flaw is that the edges in the third column are assigned to two
communities, one unique to that column, and the other used for the
third row from the top.} \label{fTGCGt216t2c3app}
\end{figure*}

\newpage
The community structure based on the partition of
3-cliques of the same benchmark graph as in figure
\ref{fTGCGt216vp}. Produced by applying the Louvain algorithm to
the $\Cthree$ clique graph, maximise modularity with $\gamma=3.0$.
Twelve communities associated with the columns are found matching
the column and row communities perfectly. This is indicated by the
edges in each row having a unique colour, and similar for the
columns.\begin{figure*}
\centering
\includegraphics[width=0.8\textwidth]{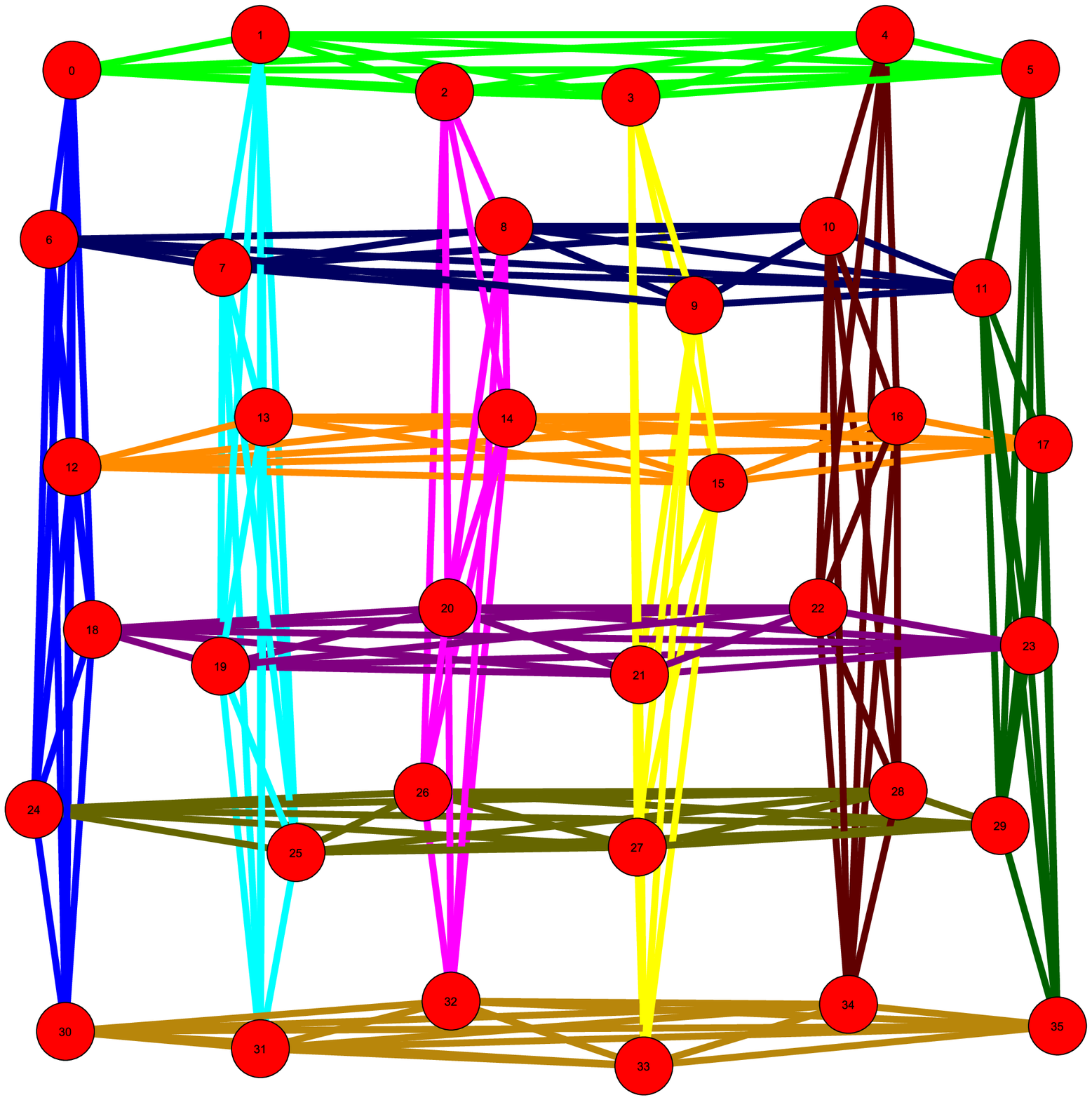}
\caption{The community structure based on the partition of
3-cliques of the same benchmark graph as in figure
\ref{fTGCGt216vp}. Produced by applying the Louvain algorithm to
the $\Cthree$ clique graph, maximise modularity with $\gamma=3.0$.
Twelve communities associated with the columns are found matching
the column and row communities perfectly. This is indicated by the
edges in each row having a unique colour, and similar for the
columns.} \label{fTGCGt72t4c3app}
\end{figure*}

\newpage
A vertex partition and an edge partition of the same
benchmark graph as used in figure \ref{fTGCGt216vp}. The edge
partition is found by finding a vertex partition of the weighted
line graph $D(G)$ described in \cite{EL09}. Both partitions are
found using the Louvain algorithm to maximise modularity
\tseref{modAdef} with $\gamma=1.2$. On this run, the vertex
partition finds the six communities associated with the rows,
indicated by the vertex shapes and colours, but the row
communities are completely missed. The edge partition find the six
columns and one of the row communities, as indicated by the edge
colours.\begin{figure}
\centering
\includegraphics[width=0.8\textwidth]{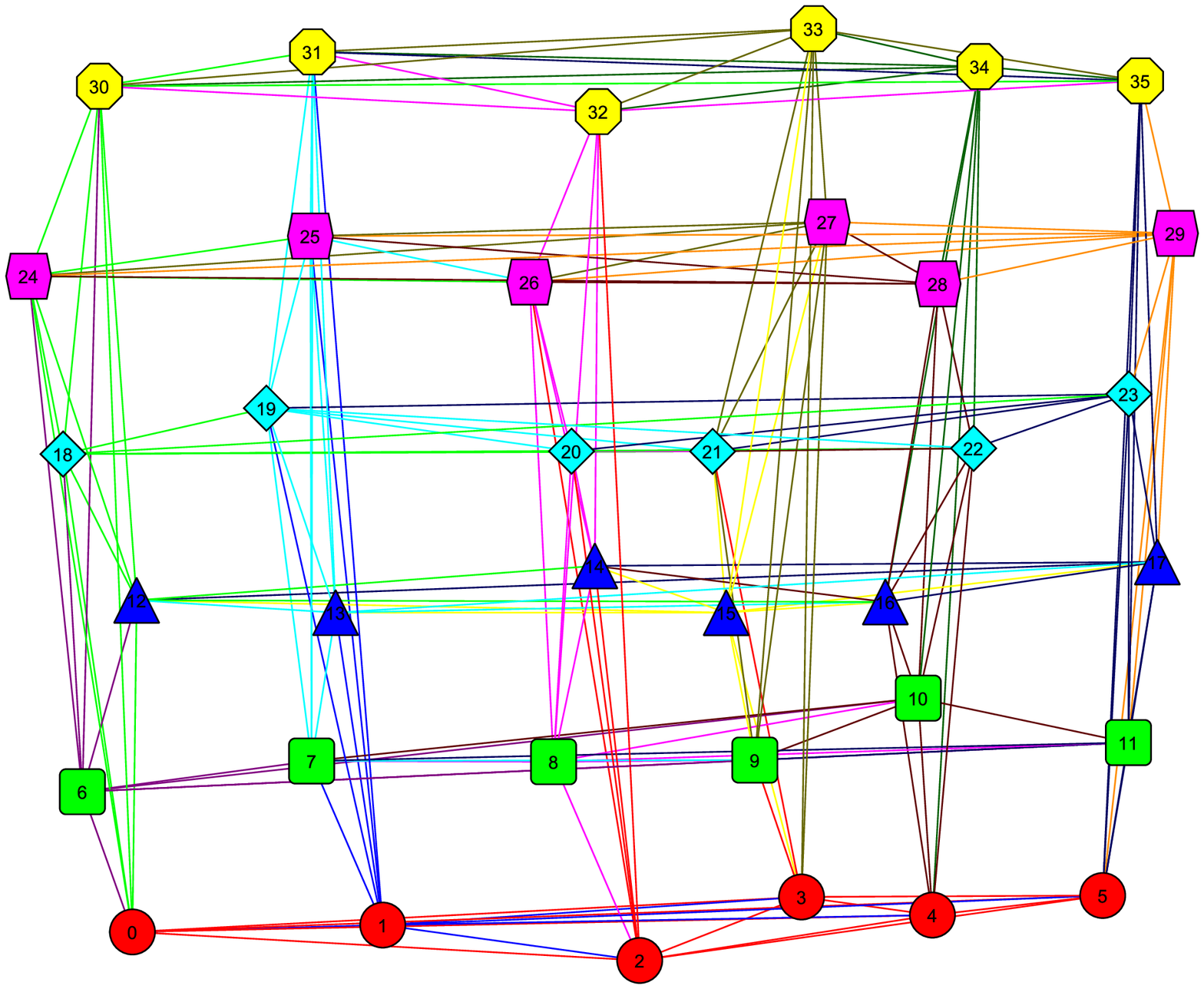}
\caption{A vertex partition and an edge partition of the same
benchmark graph as used in figure \ref{fTGCGt216vp}. The edge
partition is found by finding a vertex partition of the weighted
line graph $D(G)$ described in \cite{EL09}. Both partitions are
found using the Louvain algorithm to maximise modularity
\tseref{modAdef} with $\gamma=1.2$. On this run, the vertex
partition finds the six communities associated with the rows,
indicated by the vertex shapes and colours, but the row
communities are completely missed. The edge partition find the six
columns and one of the row communities, as indicated by the edge
colours.} \label{fTGCGt216vpep}
\end{figure}

\fi 

\if\tsecolon\tsetrue
\newpage
\section{Clique Overlap in Literature}

The abbreviations in table \ref{tcglit} on key differences in
the literature are as follows:-
\begin{itemize}
\item A superscript on the Graph Type indicates if
the set of cliques used is of fixed order (e.g. $\An$) as used
in the definitions in the text \tseref{adjA},
\tseref{adjC},\tseref{adjD}, \tseref{adjP}, or is based on the
maximal clique set, e.g.\ $\Amax$.

The notation $\Amax(G;t)$
indicates that a weighted graph $\Amax(G)$ is projected onto an
unweighted one, $\Amax(G;t)$, with all edges with weight below
(equal or above) a threshold value of $t$ being removed
(retained).
\item A superscript in a clique set $\Ccal$ indicates a limitation on the set of all cliques.
$\Ccaln$ is the set of all cliques of order $n$.
$\Ccalmaxn$ is the set of all maximal cliques of order $n$ or higher.
\item \emph{OC}  (\emph{NOC})  = (Non-)Overlapping Cliques,
\item \emph{S} (\emph{S'}) = (Simmelian) strong links;
\item \emph{SSS}=Simmelian tie (S'), all three edges strong in triad,
\item \emph{SSW}=two strong, one weak edge in triad,
\item \emph{MG}=Multigraph,
\item \emph{W} (\emph{UW}) = (un)weighted,
\item \emph{D} (\emph{UD}) = (un)directed,
\item \emph{JHC} = Johnson Hierarchical Clustering;
\item \emph{NFrat} = Newcomb's Fraternity data
\end{itemize}
\newpage

\begin{table}[htb]
\begin{tabular}{c|c|c|l}
  \hline
  \textbf{Paper} & \textbf{Clique} & \textbf{Graph} & \textbf{Other}  \\
                 & \textbf{Set}    & \textbf{Type}  & \textbf{Comments} \\ \hline\hline
  This & $\Ccaln$ & $\Cn$, $\Dn$ \\ \hline
  F92 \cite{F92} & $\Ccalthree$  & $\Amax(G;t)$ & \parbox{5cm}{$t$ chosen for NOC via S' links definition SSS/SSW accepted, $SS\emptyset$ unacceptable; threshold (t) removes $SS\emptyset$; NOC components of $S'$}\\ \hline
  F96 \cite{F96} & $\Ccalmaxn$ & $\Amaxn(G;t=1)$ & \parbox{5cm}{$n=3$; NOC via Galois Lattice; Bank Wiring Room}\\
                 & $\Ccalmaxn$ & $\Cmaxn(G;t=1)$ & \emph{ditto}\\ \hline
  F00 \cite{F00} & $\mathcal{M}$ & $A(G;\mathcal{M};t=1)$ & \parbox{5cm}{$n=3$; NOC via Galois Lattice; generalisation of \cite{F96} to general motifs $\mathcal{M}$}\\
                 &               & $C(G;\mathcal{M};t=1)$ & \emph{ditto}\\ \hline
  EB98 \cite{EB98} & $\Ccalmaxn$ & $\Amaxn(G)$ & \parbox{5cm}{NOC - JHC; Bank Wiring Room}\\
                   &             & $\Cmaxn(G)$ & \parbox{5cm}{No communities?; Bank Wiring Room}\\ \hline
  UCInet02 \cite{UCInet02} & $\Ccalmaxn$ & $\Amaxn(G)$ & \parbox{5cm}{NOC - JHC}\\
                       &             & $\Cmaxn(G)$ & \\ \hline
  PDFV05 \cite{PDFV05} & $\Ccaln$ & $\Pn(G;t=n-1)$ & UW (via threshold); OC \\ \hline
  K98 \cite{K98} & $\Ccalthree$ & $\Athree(G;t)$  & \parbox{5cm}{SSS only (= `Simmelian');  No communities; Comparison of ties;  time evolution; NFrat} \\ \hline
  K99 \cite{K99} & $\Ccalthree$    & $\Athree(G;t)$ & \parbox{5cm}{For comparison `Simmelian' (SSS) with strong ties?} \\
                 & $\Ccalmaxthree$ & $\Amaxthree$   & \parbox{5cm}{MG,UW,UD; NOC via Pearson correlation on $\sum_k \Amaxthree_{ik}\Amaxthree_{jk}$; Silicon Systems data (Firm)} \\ \hline
  KK02 \cite{KK02} & $\Ccalthree$ & $\Athree(G;t)$  & \parbox{5cm}{SSS only (= `Simmelian');  tie discussion?} \\ \hline
  KH07 \cite{KH07} & $\Ccalthree$ & $\Athree$  & \parbox{5cm}{W,D $\stackrel{\mathrm{thresh}}{\rightarrow}$ UW,UD; SSS only (= `Simmelian');  NOC via JHC on $\Athree$ (? or roles ?); tie analysis; time evolution, NFrat} \\ \hline
  PS98 \cite{PS98} & $\Ccalmaxfour$ & $\Amaxfour$ (S') &  \parbox{5cm}{Own data set, $n=4$, multigraph} \\ \hline
  B09 \cite{B09} & $\Ccalmaxfour$ & $\Amaxfour$ (S') &  \parbox{5cm}{Own data set; $n=4$; multigraph; no communities}
\end{tabular}
\label{tcglit} \caption{Table to illustrate key differences in
the literature.}
\end{table}
\newpage

\fi 

\fi 

\end{document}